\documentclass[structabstract]{aa}  
\usepackage{graphicx}
\usepackage{natbib}
\bibpunct{(}{)}{;}{a}{}{,} 
\usepackage{txfonts}
\begin{document}
   \title{Structure and Dynamics of Isolated Internetwork Ca~{\sc ii}~H Bright Points Observed by {\sc Sunrise}}

   \author{S.~Jafarzadeh\inst{1,2}, S.~K.~Solanki\inst{1,3}, A.~Feller\inst{1}, A.~Lagg\inst{1}, 
             A.~Pietarila\inst{4}, S.~Danilovic\inst{1}, 
             T.~L.~Riethm\"{u}ller\inst{1} \and V.~Mart\'{i}nez~Pillet\inst{5}
             }

   \institute{Max-Planck-Institut f\"{u}r Sonnensystemforschung, Max-Planck-Str. 2, 37191 Katlenburg- Lindau, Germany\\
              \email{shahin.jafarzadeh@mps.mpg.de}
         \and
         	 Institut f\"{u}r Astrophysik, Georg-August-Universit\"{a}t G\"{o}ttingen, Friedrich-Hund-Platz 1, 37077 G\"{o}ttingen, Germany
         \and
             School of Space Research, Kyung Hee University, Yongin, Gyeonggi 446-701, Republic of Korea
         \and
             National Solar Observatory, 950 N. Cherry Avenue, Tucson, Az 85719, USA
         \and
             Instituto de Astrof\'{i}sica de Canarias, C/Via Lactea, s/n 38205, La Laguna, E-38200, Tenerife, Spain\\             
             }

	\date{Received 24 July 2012 / Accepted 14 November 2012}

 
  \abstract
{} 
{We aim to improve our picture of the low chromosphere in the quiet-Sun internetwork by investigating the intensity, horizontal velocity, size and lifetime variations of small bright points (BPs; diameter smaller than $0.3~arcsec$) observed in the Ca~{\sc ii}~H $3968~\AA$ passband along with their magnetic field parameters, derived from photospheric magnetograms.} 
{Several high-quality time series of disc-centre, quiet-Sun observations from the Sunrise balloon-borne solar telescope, with spatial resolution of around $100~km$ on the solar surface, have been analysed to study the dynamics of BPs observed in the Ca~{\sc ii}~H passband and their dependence on the photospheric vector magnetogram signal.} 
{Parameters such as horizontal velocity, diameter, intensity and lifetime histograms of the isolated internetwork and magnetic Ca~{\sc ii}~H BPs were determined. Mean values were found to be $2.2~km~s^{-1}$, $0.2~arcsec$ ($\approx150~km$), $1.48~\langle I_{Ca} \rangle$ and $673~sec$, respectively. Interestingly, the brightness and the horizontal velocity of BPs are anti-correlated. Large excursions (pulses) in horizontal velocity, up to $15~km~s^{-1}$, are present in the trajectories of most BPs. These could excite kink waves travelling into the chromosphere and possibly the corona, which we estimate to carry an energy flux of $310~W~m^{-2}$, sufficient to heat the upper layers, although only marginally.} 
{The stable observing conditions of Sunrise and our technique for identifying and tracking BPs have allowed us to determine reliable parameters of these features in the internetwork. Thus we find, e.g., that they are considerably longer lived than previously thought. The large velocities are also reliable, and may excite kink waves. Although these wave are (marginally) energetic enough to heat the quiet corona, we expect a large additional contribution from larger magnetic elements populating the network and partly also the internetwork.}


  \authorrunning{S.~Jafarzadeh~et~al.}
   \maketitle
%

\section{Introduction}
\label{sec-intro}

The solar chromosphere is both, heavily structured and highly dynamic. One of the important contributors to the low chromosphere's structuring and dynamics in high spatial resolution images are bright points (BPs). They are also an excellent diagnostic since due to their apparent lack of internal fine structure, they are ideal tracers of horizontal motions of magnetic features. Such BPs, however, can have a rather diverse set of physical causes. Here we discuss particularly small BPs seen in the strong chromospheric lines such as Ca~{\sc ii}~H.\\
Acoustic waves, thought to be one of the heating mechanisms of the quiet chromosphere from simulations~\citep[e.g.,][]{Carlsson1997} and observations~\citep[e.g.,][]{Rutten1991,Kneer1993,Hofmann1996,Cadavid2003,Beck2008,BelloGonzlez2010} appear as bright, short-lived points in Ca~{\sc ii}~H/K filtergrams that are sufficiently spectrally narrow to isolate the H$_{2V}$/K$_{2V}$ emission peak. These bright points are normally referred to as H$_{2V}$/K$_{2V}$ grains. With the help of radiation-hydrodynamic numerical simulations~\citet{Carlsson1997} explained the origin of these point-like brigthenings as high-frequency $p$-modes.~\citet{Lites1993} found no clear correlation between the Stokes $V$ signal and the occurrence of K$_{2V}$ grains, observed in the Ca~{\sc ii}~H passband, in contrast to~\citet{Sivaraman1982}, who had concluded that these grains are associated with a weak magnetic field of $20~G$ on average.~\citet{Cadavid2003} found that most K-line brightenings occur 2 minutes before or after $G$-band darkening events in roughly the same positions in an internetwork area. They concluded that the timing and coupling of these chromospheric brightenings and photospheric darkenings are due to a pre-existing acoustic wave pattern with four minute periodicity. They also showed that the K$_{2V}$ grains are not associated with any detectable magnetic flux in the simultaneously observed magnetograms. The time delay of about two minutes is roughly consistent with the travel time of sound from the photosphere to the chromosphere~\citep{Hoekzema2002}.\\
In addition to the short-lived acoustic H$_{2V}$/K$_{2V}$ grains, other bright, point-like structures are observed in the low solar chromosphere and around the temperature minimum\footnote[1]{Note that the term ``bright point'' has been interchangeably used in the literature for chromospheric phenomena such as brightenings associated with magnetic features, or K-grains~\citep[e.g.,][]{Steffens1996}, and for bright, point-like features in the other layers of the solar atmosphere, e.g., X-ray bright points in the corona, or photospheric magnetic bright points. Here we restrict ourselves to the chromospheric features.}. These include the brightest parts of the (1) ``reversed granulation''~\citep{Rutten2004}, which appears as bright arcs all over internetwork areas, concentrated to brighter point-like structures in some regions, (2) magnetic bright, point-like features in both network and internetwork areas, and (3) ``persistent flashers'', i.e., BPs which appear (when intermittent magnetic patches are squeezed together) and disappear (when the magnetic concentrations become less dense; as imposed by the granular flows) during their lifetime~\citep{Brandt1992,de-Wijn2005,Rutten2008}.\\
Earlier studies of BPs in the quiet Sun mostly concentrated on the dynamics of network BPs. These have been well-studied at the atmospheric layers we sample, i.e., from the upper photosphere to the low chromosphere~\citep[e.g.,][]{Lites1993,Wellstein1998}. They are connected to photospheric BPs~\citep[e.g.,][]{Riethmuller2010} with concentrated kilo-Gauss, nearly vertical magnetic elements described by flux tubes~\citep[e.g.,][]{Spruit1976,Solanki1993,Yelles-Chaouche2009}, and have been often referred to as MBPs (magnetic bright points; e.g., see~\citealt{Muller2000,Mostl2006}) or GBPs (when identified in the $G$-band passband; e.g.,~\citealt{Berger2001,Steiner2001,Sanchez-Almeida2001,Sanchez-Almeida2010,Schussler2003,Shelyag2004,Bovelet2008}).\\
The chromospheric magnetic internetwork BPs are more challenging and less studied. The few previous investigations of magnetic internetwork Ca~{\sc ii}~H BP dynamics~\citep[e.g.,][]{Al2002,de-Wijn2005}, although of considerable value, suffer from the comparatively limited spatial resolution of the employed data. A complete observational picture of the dynamics of these bright point-like features obtained at high spatial resolution, including the intensity, horizontal velocity, size and lifetime distributions, is so far unavailable. Providing such a picture for the smallest elements currently observable is the aim of this paper.\\ 
Here we characterize the time-dependent properties, including the dynamics of the magnetic BPs observed in the low chromospheric layers sampled by the {\sc Sunrise} Filter Imager (SuFI) Ca~{\sc ii}~H filter, centred at $3968~\AA$. We measure their magnetic property with the help of simultaneous full Stokes photospheric observations in the Imaging Magnetograph eXperiment (IMaX) Fe I $5250.2~\AA$ passband. Several criteria are considered to define the particular breed of Ca~{\sc ii}~H BPs probed in this work.\\
This paper is organised as follows: in Sect. 2, we describe the data and their preparation prior to analysis. In Sect. 3, our methods for studying the temporal evolution of the identified {\sc Sunrise} Ca~{\sc ii}~H BPs, as well as their magnetic correspondence measurements are explained. The results of our investigations and related discussions are presented in Sect. 4, followed by the conclusions in Sect. 5.
\begin{table*}[h!t]
\caption{Log of observations.}      
\label{table:obslog}   
\centering                
\begin{tabular}{l c c c | c c l c c | c c | c c l c c | c c}  
\hline\hline        
Central & FWHM & \multicolumn{4}{c}{9 June 2009} && \multicolumn{6}{c}{11 June 2009} && \multicolumn{4}{c}{13 June 2009} \\
\cline{3-6} \cline{8-13} \cline{15-16} \cline{17-18}
Wave- & of Filter & \multicolumn{2}{c}{00:36-00:59$^{\mathrm{ *}}$} & \multicolumn{2}{c}{01:32-02:00} && \multicolumn{2}{c}{15:22-15:44} & \multicolumn{2}{c}{20:09-20:19} & \multicolumn{2}{c}{21:00-21:09} && \multicolumn{2}{c}{01:31-01:35} & \multicolumn{2}{c}{01:46-01:59}\\ 
\cline{3-4} \cline{4-6} \cline{8-9} \cline{10-11} \cline{12-13} \cline{15-16} \cline{17-18}
length &  & Exp.$^{\mathrm{ **}}$ & Cad.$^{\mathrm{ ***}}$ & Exp. & Cad. && Exp. & Cad. & Exp. & Cad. & Exp. & Cad. && Exp. & Cad. & Exp. & Cad.\\
(\AA) & (\AA) & (ms) & (s) & (ms) & (s) && (ms) & (s) & (ms) & (s) & (ms) & (s) && (ms) & (s) & (ms) & (s)\\   
\hline                   
   3000 & 50 & 500 & 12 & 500 & 12 && - & - & 130 & 8 & 130 & 7 && - & - & - & -\\  
   3120 & 12 & 150 & 12 & 150 & 12 && - & - & 250 & 8 & 250 & 7 && - & - & - & -\\
   3880 & 8 & 80 & 12 & 80 & 12 && 65 & 4 & 70 & 8 & 70 & 7 && 75 & 3 & 75 & 3\\
   3968 & 1.8 & 960 & 12 & 960 & 12 && 500 & 4 & 500 & 8 & 500 & 7 && 500 & 3 & 500 & 3\\
   5250.2 & 0.085 & 250 & 33 & 250 & 33 && - & - & - & - & - & - && 250 & 33 & 250 & 33\\
\hline                        
\end{tabular}
\begin{list}{}{}
\item[$^{\mathrm{*}}$] All times in UT
\item[$^{\mathrm{**}}$] Exposure Time
\item[$^{\mathrm{***}}$] Cadence
\end{list}
\end{table*}

\section{Observations and Data Preparation}
\label{sec-obs}

Several simultaneous image sequences obtained by the {\sc Sunrise} Filter Imager (SuFI;~\citealt{Gandorfer2011}) and the Imaging Magnetograph eXperiment (IMaX;~\citealt{Martinez-Pillet2011}) on board the {\sc Sunrise} balloon-borne solar telescope~\citep{Solanki2010,Barthol2011,Berkefeld2011} are used for this study.\\
Table~\ref{table:obslog} summarises the SuFI and IMaX observations used in this study. In this table, we present exposure time and observing cadence of each image sequence and for all wavelengths recorded by {\sc Sunrise}. However, only the SuFI Ca~{\sc ii}~H (centred at $3968~\AA$), SuFI CN (centred at $3880~\AA$) and IMaX Fe I $5250.2~\AA$ images were used for this paper. Image sequences that were not observed in a certain wavelength and/or data that were not available (e.g., not yet reduced) at the time of this investigation, are indicated by dashed lines.

\subsection{Ca~{\sc ii}~H Images}
\label{subsec-CaIIHimages}

The primary time series in this investigation consist of SuFI images~\citep{Hirzberger2010} taken in the Ca~{\sc ii}~H filter (centred at $3968~\AA$ with FWHM$\approx 1.8~\AA$). The series were recorded on 2009 June 9 (at $0$ and $1$ UT, both with $12~sec$ cadence), June 11 (at $15$, $20$ and $21$ UT, with $4$, $8$ and $7~sec$ cadences, respectively) and June 13 (at $1$ UT, with $3~sec$ cadence) in quiet-Sun regions close to disc-centre. We employ level $3$ data, i.e., images that have been phase diversity reconstructed with averaged wave front errors (cf.~\citealt{Hirzberger2010,Hirzberger2011}).\\
To obtain an estimate of the formation height of the radiation passing through the {\sc Sunrise}/SuFI Ca~{\sc ii}~H passband, we determined the contribution functions (CF) by means of the 1D RH-code~\citep{Uitenbroek2001}. This code solves both radiative transfer and statistical equilibrium equations in non-LTE for a given atmospheric model. We used a four-level plus continuum Ca~{\sc ii}~H model atom~\citep{Uitenbroek2001} and three different atmospheric models, FALC, FALF and FALP, which respectively describe an averaged quiet-Sun area, bright regions of the quiet-Sun and a typical plage region~\citep{Fontenla1993,Fontenla2006}. For all atmospheric models, the intensity profiles (synthetic spectra) of the Ca~{\sc ii}~H line, computed with the RH-code from the atomic model, as well as the determined contribution functions to line depression~\citep[e.g.,][]{GrossmannDoerth1988} were convolved with the transmission profile of the {\sc Sunrise}/SuFI Ca~{\sc ii}~H filter. Fig.~\ref{fig:fh}a shows the Ca~{\sc ii}~H spectra before convolution with the transmission profile (dashed curve) for FALC (red line), FALF (green line) and FALP (blue line). Fig.~\ref{fig:fh}b displays the convolved profiles. The latter model represents higher H$_{2V}$/H$_{2R}$ emission peaks around the line-core compared to the other models due to excess heating related to the magnetic field, just as in magnetic BPs~\citep[e.g.,][]{Skumanich1975,Ayres1986,Solanki1991}. For a comparison of Ca~{\sc ii}~K line shapes in different model atmospheres see Fig.~$7$ of~\citet{Fontenla2009}. The mean intensity contrast of FALP relative to FALC ($\approx1.43$; determined from the convolved spectra) is similar to that of small, internetwork {\sc Sunrise} Ca~{\sc ii}~H BPs under study (see Sect.~\ref{subsubsec-int}). Integration of the (filter-profile convolved) CF over wavelength gives the contribution of each height to formation of the signal seen in the {\sc Sunrise} Ca~{\sc ii}~H bandpass. Plotted in Fig.~\ref{fig:fh}c are the computed CFs for the {\sc Sunrise} filter versus geometric height for the three model atmospheres. The CFs have multiple peaks and cover a wide range of heights from below the surface to the mid-chromosphere. The vertical dotted lines in Fig.~\ref{fig:fh}c indicate the average formation heights~\citep[cf.,][]{Leenaarts2012}, lying at $437$, $456$ and $500~km$ for FALC, FALF and FALP, respectively. The sudden drop of the CF's curve for FALP at $\approx1700~km$ is due to the rapid increase in the temperature at the top of the chromosphere.\\
The {\sc Sunrise}/SuFI Ca~{\sc ii}~H filter is narrower than that of \textit{Hinode} (centred at $3968.5~\AA$ with FWHM$\approx 3.0~\AA$;~\citealt{Tsuneta2008}), so it samples higher layers in the quiet-Sun compared to the \textit{Hinode} Ca~{\sc ii}~H passband with an average formation height of $247~km$.~\citep[cf.][]{Carlsson2007}.\\
The photon noise is the most prominent noise source in intensity and hence it is assumed as the only noise contribution in the measurements. The noise introduced by instrumental effects into the {\sc Sunrise} observations is much smaller and not considered further. The photon noise is calculated by measuring the high frequencies derived from the Fourier power spectrum of the reconstructed images (after apodization). A value of $0.01~\langle I_{Ca} \rangle$ was measured as the averaged photon noise in our Ca~{\sc ii}~H images.
\begin{figure*}[tbp]
	\centering
	\includegraphics{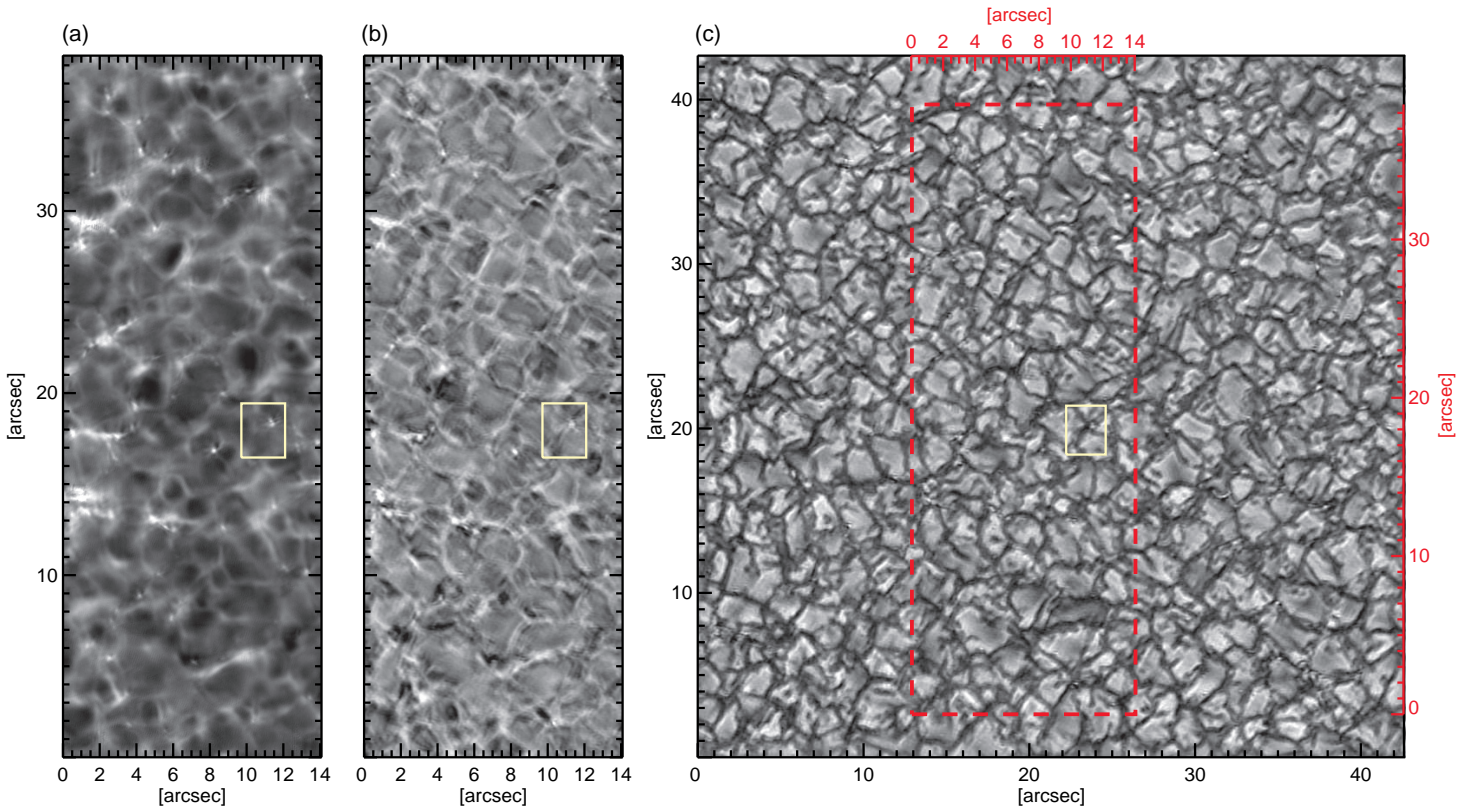}
	\caption{Example of co-spatial and co-temporal images from both SuFI and IMaX instruments on board the {\sc Sunrise} balloon-borne solar telescope. ($a$) SuFI Ca~{\sc ii}~H image. ($b$) Average of two line-positions at $-40~m\AA$ and $+40~m\AA$ from the Fe I $5250.2~\AA$ line-core. Plotted is the part of the full FOV overlapping with the SuFI Ca~{\sc ii}~H image. ($c$) Full field of view of the continuum position of the IMaX Fe I $5250.2~\AA$ line. The red dashed-line in this image illustrates the co-aligned area in common with the SuFI Ca~{\sc ii}~H image. The yellow box (solid-line) includes a sample Ca~{\sc ii}~H BP studied here; its temporal evolution and parameter fluctuations are presented in Fig.~\ref{fig:bp_temp} and Fig.~\ref{fig:bp_var}, respectively. The common coordinate system adopted for the present analysis is given below and to the left of panels $a$ and $b$, as well as above and to the right of panel $c$.}
	\label{fig:obs_examples}
\end{figure*}

\begin{figure}[tbp]
	\centering
	\includegraphics[width=8cm]{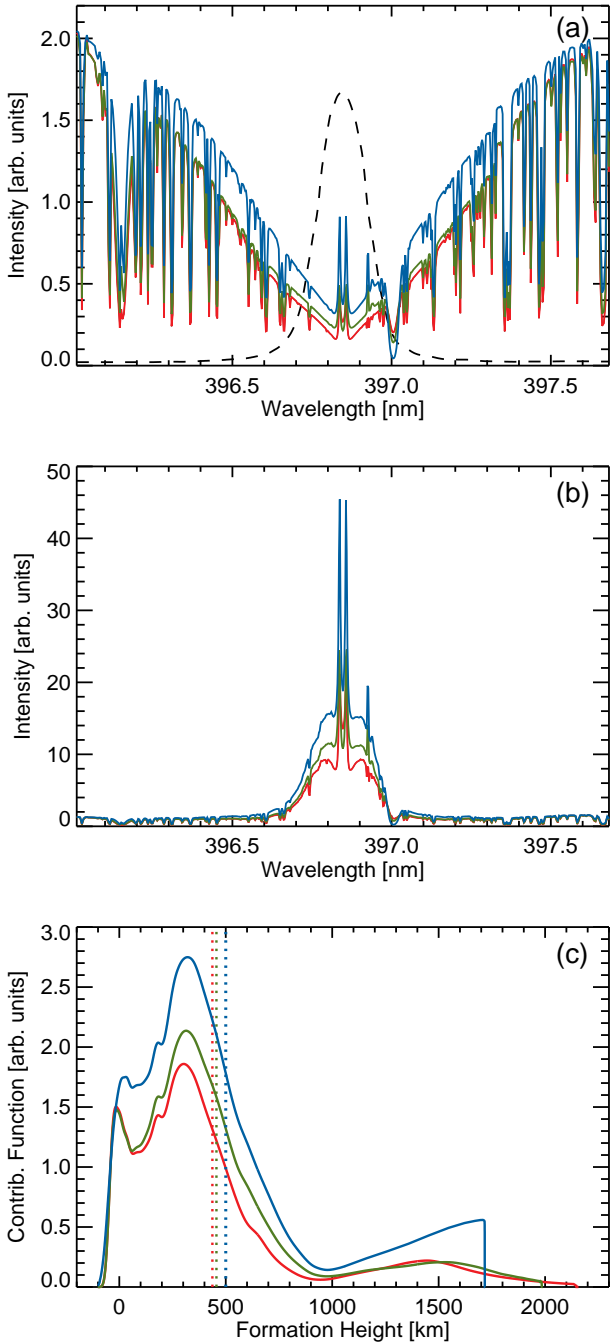}
	\caption{($a$): Intensity profiles (synthetic spectra) of Ca~{\sc ii}~H line for three atmospheric models (see text). The dashed curve represents the transmission profile of the {\sc Sunrise}/SuFI Ca~{\sc ii}~H filter.
	  ($b$): Results of the convolution of the spectra with the transmission profile shown in panel ($a$).
	  ($c$): Line depression contribution functions for the {\sc Sunrise}/SuFI Ca~{\sc ii}~H bandpass vs. height. The vertical lines indicate the corresponding average heights formation.
Red, green and blue lines in all panels represent the FALC, FALF and FALP model atmospheres.}
	\label{fig:fh}
\end{figure}

\subsection{Stokes Measurements}
\label{subsec-stokes}

Full Stokes vector ($I$,$Q$,$U$,$V$) data recorded in parallel to the Ca~{\sc ii}~H sequences by IMaX in the magnetically sensitive photospheric Fe I $5250.2~\AA$ line at a cadence of $33~sec$ were also analysed whenever available. The Stokes vector was measured in four wavelength positions in the line located at $-80$, $-40$, $+40$ and $+80~m\AA$ from line centre and at $+227~m\AA$ in the continuum. The noise level was $3\times10^{-3}~I_{c}$ after phase-diversity reconstruction ($V5-6$ data ;~\citealt{Martinez-Pillet2011}).\\ 
In order to study the magnetic properties of the investigated BPs, following \citet{Solanki2010} and \citet{Martinez-Pillet2011}, we averaged over Stokes $V$ at the $4$ wavelength positions inside the line, normalised to the local quiet-Sun continuum intensity ($I_{c}$). To avoid cancellation effects, the sign of the two red wavelength points was reversed prior to averaging. This wavelength averaged Stokes $V$ will be called $CP$ in the following.\\
The total linear polarisation ($LP$) was also computed from the Stokes $Q$ and $U$ profiles ($LP=\sqrt{Q^2+U^2}$), averaged over the four wavelength positions in the line. To do this, we first average the Stokes $Q$ and $U$ profiles over the wavelength positions with the same sign (i.e., the two outermost and two innermost wavelength positions), then compute the $LP$ for each averaged position, and finally calculate the mean value of the two $LP$s. The $CP$ and $LP$ were determined from the non-reconstructed data (with a factor of $2$ lower spatial resolution and, more importantly, a factor of $3$ lower noise compared to the reconstructed data), which were flat-fielded and corrected for instrumental effects.\\
The noise level for $LP$ was computed from the IMaX Stokes $Q$ and $U$ frames of the continuum wavelength and the $LP$ was corrected for the offset produced due to squaring the Stokes profiles. Furthermore, to increase the signal-to-noise ratio ($SNR$), an average over a $3\times3$ pixels area centred at the corresponding location of the BP on the aligned IMaX images was performed on stokes $Q$, $U$ and $V$ profiles, at each wavelength position, prior to computing the $CP$ and $LP$ signals. The $3\sigma$ level of the Stokes profiles decreases due to the averaging over four wavelength positions and it is additionally reduced due to the spatial averaging of the polarisation signals. The reduction in the noise by a factor of $3$ due to the spatial smoothing over $3\times3$ pixels was not fully achieved due to coherent noise, probably caused by image jitter. The final computed value of the $3\sigma$ noise levels for $CP$ and $LP$ normalised to $I_{c}$ is on average $1\times10^{-3}$.\\
Areas where the $CP$ signal is greater than the computed $3\sigma$ noise level are defined as magnetic regions (patches).

\subsection{Image Alignment}

Since the Ca~{\sc ii}~H images form the primary data set considered in this paper, the IMaX $50\arcsec\times50\arcsec$ images ($936\times936$ pixels in size and with an image scale of $0.0523~arcsec/pixel$) were cropped to the field of view (FOV) of the SuFI Ca~{\sc ii}~H images covering $712\times1972$ pixels (with an image scale of $0.0207~arcsec/pixel$) after interpolating the images to an approximately common scale and orientation.\\
Furthermore, the images were properly aligned by applying a cross-correlation technique on common sets of manually selected features in two image series having the closest formation heights: Ca~{\sc ii}~H and the average of the $-40$ and $+40~m\AA$ line-positions of the IMaX Stokes $I$. The latter corresponds to the inner flanks of Fe I $5250.2~\AA$ and is a proxy for the line-core intensity. The differences in solar position of the manually selected features were iteratively minimized using the Powel minimization function. The resulting cross-aligned images show good alignment within the spatial resolution of the observations. After co-alignment, the IMaX images were given the same coordinate system as the SuFI Ca~{\sc ii}~H images for each image sequence, so that any spatial structure, e.g., a BP, is addressable by the same coordinates in all images. Note that the SuFI and IMaX data have different cadences, hence, we selected those Ca~{\sc ii}~H images whose observing times are closest to the IMaX line-core observations.\\
Fig.~\ref{fig:obs_examples} shows example frames of SuFI Ca~{\sc ii}~H (left), IMaX line-core (as described above; centre) and the IMaX Stokes $I$ continuum (right) after the co-alignment.

\section{Analysis}
\label{sec-analysis}

In Sect.~\ref{sec-intro}, we reviewed point-like brightenings observed around the temperature minimum, i.e., H$_{2V}$/K$_{2V}$ grains, product of reversed granulation, magnetic BPs and persistent flashers. They are in many respects similar to the BPs considered here, but many of the BPs in the literature also differ in important ways. In the present work, we investigate the {\sc Sunrise}/SuFI Ca~{\sc ii}~H BPs which ($1$) are located in internetwork areas, ($2$) are intrinsically magnetic (as far as we can determine), ($3$) are not oscillations or wave-like features, ($4$) are not the product of reversed granulation, ($5$) are actual bright point-like features, i.e., have a roughly circular shape, ($6$) are brighter than the mean intensity of the image and ($7$) do not interact with other BPs (i.e., merge or split) in their observed lifetimes. These conditions restrict us to internetwork, magnetic, non-interacting and small BPs seen in the height sampled by the {\sc Sunrise}/SuFI Ca~{\sc ii}~H passband. These BPs turn out to be the smallest Ca~{\sc ii}~H BPs observed so far. The small size is partly due to the upper limit of $0.3~arcsec$ that we set. This facilitates precise tracking and avoids any ambiguities due to a changing shape and/or varying distribution of the brightness over the magnetic internetwork BPs. In order to isolate BPs with the above properties, we set the following criteria: 

\begin{enumerate}
\item [(1)] All the BPs are investigated in quiet internetwork areas where no significant large bright features, i.e., no obvious magnetic structure representing the network, are present in the immediate vicinity, i.e., within $3~arcsec$. 
\item [(2)] The magnetic properties of the BPs are determined from the simultaneous photospheric magnetograms (Stokes $V$ signals) obtained from the {\sc Sunrise} IMaX instrument. Only BPs with $CP$ above $3\sigma$ during at least half of the BPs' lifetime are considered. This criterion is imposed whenever simultaneous IMaX observations are available (see Table~\ref{table:obslog}).
\item [(3)] Criterion ($2$) ensures that the analysed BPs are not due to non-magnetic phenomena such as acoustic waves. Since the magnetic information is not available for the image sequences observed on 11 June 2009, we perform a further test for these time series proposed by~\citet{Cadavid2003}. They found a delay between photospheric intergranular darkenings and $H$/$K$-line brightenings. We investigate such an occurrence for our candidate BPs by comparing the aligned simultaneous SuFI images in the CN $3880~\AA$ and Ca~{\sc ii}~H passbands. Thus, a BP is included only if its brightening is not preceded by a CN darkening.
\item [(4)] The bright arc-shaped features caused by the reversed granulation may in some cases produce point-like brigthenings and hence may lead to false BP detection. These arcs and associated BPs are, however, short-lived. Therefore, in order to avoid misidentification, we exclude BPs with lifetimes less than $80$ seconds~\citep{de-Wijn2005}.
\item [(5)] We set an upper limit of  $0.3~arcsec$, i.e $2$ times the spatial resolution of the {\sc Sunrise} data, for the BP diameter. This size threshold limits the identified bright features to actual point-like features with roughly circular shapes.
\item [(6)] A BP is considered to be detected only if its intensity (i.e., contrast) in a Ca~{\sc ii}~H image, at any given time, rises above the mean intensity of the entire frame ($\langle I_{Ca} \rangle$).
\item [(7)] Finally, BPs which merge with other features or split into several BPs (or magnetic elements) during their lifetimes are excluded. In order to sort out BPs affected by such interactions an additional careful manual inspection is made of the temporal evolution of each BP.
\end{enumerate}

\subsection{Tracking Algorithm}
\label{subsec-tracking}

Identification and tracking of  Ca~{\sc ii}~H BPs, i.e., determining their positions and other parameters in consecutive frames, is not straight forward as pointed out by~\citet{de-Wijn2005}. We employ a tracking algorithm based on a particle tracking approach developed by~\citet{Crocker1996}. This method has been widely used and has been confirmed as a precise approach in colloidal studies for identifying and tracking small roughly point-like features on a noisy and variable background~\citep[e.g.,][]{Crocker1996,Weeks2000,Jenkins2008}, conditions very similar to the Ca~{\sc ii}~H images.\\
After the BPs were automatically identified in all frames, we again visually checked each BP in each image in order to catch and discard misidentifications, i.e., the interacting BPs. Then, starting from the frame in which a particular BP was first identified we tracked the selected BP automatically backward and forward in time for as long as all criteria listed above were met.\\
The brightness, size and shape of a BP change as it migrates. The tracking algorithm must deal with changing properties and distinguish the BP from noise and other transient bright, point-like features (e.g., caused by a passing wave) in the immediate surroundings of the candidate BP.\\
A region slightly larger than the area that the selected BP is moving in (i.e., a typical area of $3^{\prime\prime}\times3^{\prime\prime}$) is cropped from the whole field of view. This smaller subframe generally includes fewer non-interesting bright features that may cause difficulties to the tracking procedure.\\
The algorithm is then initialized by determining all local intensity minima and maxima in the first subframe (the reference frame), followed by two steps applied to all images of the whole time series: (1) image restoration and (2) locating the BP and linking the locations into trajectories.

\subsubsection{Image Processing}
\label{subsec-restoration}

\begin{figure}[thp]
	\centering
	\includegraphics[width=5cm]{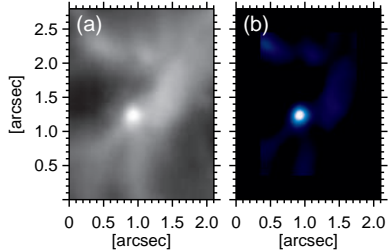}
	\caption{A cropped subframe including a visually selected BP (panel $a$) is restored after its background is subtracted off and noise is reduced (panel $b$).
              }
	\label{fig:bp_track}
\end{figure}

In order to facilitate precise tracking of the identified BPs, the image is restored to correct for inhomogeneities such as non-uniform background solar intensity, noise and geometric distortion due to rectangular pixels. The restoration includes a correction for sudden brightenings in the immediate surroundings of the BP, e.g., caused by a passing wave. This effect includes local intensity peaks of varying levels which in principle cannot be distinguished from the BP even by applying an intensity threshold. In this study, we follow a sophisticated method based on the algorithm developed by~\citet{Crocker1996} to track suspended particles. The algorithm restores the image by applying a convolution kernel in which ($1$) the low-frequency modulation of the background intensity with a non-uniform brightness is subtracted from the entire image after boxcar averaging over a circular region with a diameter of $2w+1$ pixel, where $2w$ is an even valued integer slightly larger than a single BP's apparent diameter in pixels, and ($2$) purely random noise with a correlation length of $\lambda_{n} \approx 1$ (i.e., assuming single pixel instrumental noise;~\citealt{Jenkins2008}) is suppressed without blurring the image. The latter step is performed by convolving each image with a Gaussian kernel. In order to maintain consistency with our BP definition, we set the former size threshold $2w$ slightly larger than the maximum diameter of the defined BPs (i.e., $0.3~arcsec$) which equals $20$ pixels. Assuming a fixed size of $2w$ for the BPs, we compute both these steps with the convolution kernel

\begin{equation}
	K(i,j)= \frac{1}{K_{0}} \left [ \frac{1}{B} \exp \left ( -\frac{i^{2}+j^{2}}{4 {\lambda_{n}}^{2}} \right ) -\frac{1}{\left ( 2w+1 \right )^{2}}\right ]\,,
	\label{equ:kernel}
\end{equation}

\noindent
where $K_{0}$ and $B$ are normalisation constants. Thus, the filtered image ($A_{f}$) after the restoration is given by

\begin{equation}
A_{f}(x,y) = \sum_{i,j=-w}^{w}A(x-i,y-j) K(i,j)\,,
	\label{equ:restored_image}
\end{equation}

\noindent
where $A$($x$, $y$) is the original image, ($x$, $y$) and ($i$, $j$) are pixel coordinates in the image and kernel, respectively.
As an example, panel ($b$) in Fig.~\ref{fig:bp_track} shows the result of such an image restoration applied to the original image shown in panel ($a$).~\citet{Berger1995} restored $G$-band images by applying a somewhat similar kernel from a ``blob finding'' algorithm used to identify GBPs, followed by a further intensity enhancement of dilation and erosion processing.\\
To summarise, we applied a real-space, spatial bandpass filter by (wavelet) convolution of the Ca~{\sc ii}~H images with  ($1$) a ``large kernel'' to cut off low spatial frequencies (i.e., eliminating the larger scale residual noise of the bright, extended background structures; the second term in Eq.~(\ref{equ:kernel})), and ($2$) a ``small kernel'' to truncate high frequency noise (the first term in Eq.~(\ref{equ:kernel})); while retaining information of a characteristic size ($2w+1$). As a result, we enhanced the BP's intensity profiles at the expense of other image components, e.g., passing waves and reversed granulation. However, the enhanced images were only used to locate the BPs and to measure their sizes. They were not used in subsequent measurements of the BP intensity (i.e., contrast), since the restoration procedure may introduce spurious high spatial frequency artifacts in intensity. The contrast values of the BPs are instead measured from the original Ca~{\sc ii}~H images, by referencing their accurate positions and sizes obtained from the restored images.

\subsubsection{BP's Location and its Trajectory}
\label{subsubsec-location}

The position of the maximum intensity in the identified (selected) BP in the reference frame is assigned the initial coordinates of the BP. These initial coordinates are then assumed to be at the centre of a circular window with a fixed diameter of $20$ pixels $\approx0.4~arcsec$ (i.e., a window slightly larger than the maximum allowed size of the BPs with diameter of $0.3~arcsec$). If more than one pixel has the same maximum intensity value, the pixel halfway between them is selected instead. Then, ($1$) the centre of gravity of the intensity is computed for pixels inside the circular window as new coordinates of the BP, ($2$) the circular centroid is moved to the new coordinates and ($3$) the coordinates are iteratively checked and refined if the new centre is found to have a larger offset than $0.5$ pixel from the current circular window centroid.\\
In the next step, the selected BP in the reference frame is compared with the automatically identified BP in the vicinity of its position in both the preceding and the following frames. Finally, the locations of a BP in all consecutive frames are linked together to describe its trajectory.\\
The displacement of a BP between consecutive already analysed frames can be used as an input in the tracking algorithm and is useful when a BP disappears and re-appears as happens for the so-called ``persistent flashers". This allows the missing points to be interpolated in the trajectory. Fig.~\ref{fig:bp_temp} illustrates an example of a time series for a selected BP and its corresponding trajectory as computed by the tracking algorithm. Time proceeds row by row and from left to right.\\
Normally, the standard image processing and tracking algorithms locate a BP with an uncertainty of $1/N$ of a pixel~\citep{Hunter2011a}, where $N$ is the diameter of the BP in pixels. \citet{Crocker1996} did an error analysis for the tracking technique employed here and came to the conclusion that the error for locating the objects (BPs in our case) can have a minimum value somewhat better than $0.05$ pixel. However, considering the fact that the size and intensity variation of such BPs may affect this precision~\citep{Hunter2011b}, we (over-) estimate that our algorithm is able to locate the BPs within half a pixel~\citep{Crocker1996}. 

\subsection{Properties of the BPs}

For each BP in every frame, in addition to the location, a number of further parameters were determined, as described below. Some quantities, such as intensity and size (diameter), are dependent upon the threshold used to identify the BP or compute these parameters, since the threshold defines which pixels are included in the measurements. This point should be kept in mind when the results obtained in the following are compared with the other, previously found values. The difference in the spatial resolutions of different observations, however, is also considerable and very likely has a significantly stronger effect.\\
A BP is considered to be detected if its intensity (i.e., contrast) in a Ca~{\sc ii}~H image, at any given time, rises above the mean intensity of the entire frame ($\langle I_{Ca} \rangle$). Although this criterion may cause the relatively low contrast BPs (i.e., the ones which are less bright than the mean intensity of the whole image, but have locally high contrast) to be missed, it reduces the chance of false detections.\\
We discard interacting BPs to avoid falsely assigning different features to the same BP in the course of its lifetime. This criterion may bias the lifetime distribution to shorter lifetimes since the longer lived BPs are statistically more likely to interact with other features. In addition, limiting ourselves to non-interacting BPs may introduce a bias in the proper motion speeds, although this is not clear. Thus, such non-interacting BPs may move slower since the chances for interaction increases for the faster moving BPs that travel a greater distance. Conversely, the process of interaction could slow the BPs down, so that the selected BPs may be biased towards higher proper motions.\\
The reader should keep all such biases (caused by the selection criteria) in mind when the properties of our selected BPs are compared with previous findings (where different selection criteria were applied and which therefore suffer from different biases).

\subsubsection{Size}
\label{subsubsec-size}

We apply two methods to measure the BP sizes.\\
\textbf{Method ($1$)}:~The size (diameter) of a BP is given by the radius of a circle with the same area as the BP. Following~\citet{Crocker1996}, the effective radius $R$ of a {\sc Sunrise} Ca~{\sc ii}~H BP with arbitrary shape is calculated as the root mean square (rms) of intensity-weighted distances of all pixels within a BP from its location i.e., its centre of gravity of intensity obtained as in Sect.~\ref{subsubsec-location}. This calculation is carried out within the circular window defined in Sect.~\ref{subsubsec-location}, which, by construction, is always larger than the BP (and the BP's intensity will have dropped to below the threshold in all directions within this circle). This radius is misleadingly called ``Radius of Gyration" in the particle tracking algorithm. It is misleading because it is not related to gyration or rotation of a particle or a BP. Given its success in following point-like sources on a noisy and variable background, we believe that this is an appropriate method for measuring the size of Ca~{\sc ii}~H BPs since it remains largely unaffected by a sudden brightening due to, e.g., a passing wave, etc. Hence, it is our method of choice.\\
\textbf{Method ($2$)}:~Sizes of BPs are estimated based on the Full Width at Half Maximum ($FWHM$) of the intensity peak at the location of the BP.\\
The $FWHM$ of a 2D Gaussian fit to the intensity profile of photospheric BPs has been widely used for estimating their sizes~\citep[e.g.,][]{Title1996, Sanchez-Almeida2004}. The 2D Gaussian functions must deal with the narrow and dark intergranular lanes in which the photospheric BPs typically reside. Unlike the photospheric BPs, the Ca~{\sc ii}~H BPs are embedded in a relatively dark internetwork area with passing bright waves. In addition, the higher contrast of Ca II H images compared to photospheric passbands (e.g., $G$-band) as well as the fact that the Ca~{\sc ii}~H data sample higher layers of the solar atmosphere, imply a larger size and more shape variation of the detected BPs. In particular, the shape variations decrease the quality of the fits with any specific 2D Gaussian function to the identified Ca II H BPs. An alternative approach is to measure the area across the intensity profile of the BP when it drops to half of its intensity peak (maximum) value. Assuming a roughly circular shape of the BPs and hence a circular level set for the Gaussian distribution, the (mean) $FWHM$ of a BP is calculated as the diameter of a circle with the same area as measured.\\
The latter definition (interpretation with a $FWHM$) results in diameter that are on average $21$\% larger and more variable than the ones obtained from the tracking algorithm. We prefer the definition based on the tracking algorithm (\textit{method} $1$), since it deals better with sudden brightenings of the image when travelling bright waves cross the BP. At such moments \textit{method} $2$ can give an arbitrarily large size. Therefore, the size values used in this paper are based on the first definition.\\

\subsubsection{Lifetime}
\label{subsubsec-lifetime}

The birth and death times of many BPs were not observable in our image sequences mainly because of the relatively short time series as well as the relatively small field-of-view. Therefore, the lifetime was computed only for those BPs that were born and died within the course of the time series and which did not enter or leave the field-of-view during the time series. A BP is considered to be born when its brightness first rises above $\langle I_{Ca} \rangle$ and is considered dead when its intensity is observed to permanently drop below $\langle I_{Ca} \rangle$, i.e., it does not rise again to above this detecting threshold for $160~sec$ (i.e., $2$ times our lifetime threshold) in the following frames. These additional frames are considered in order to avoid underestimating the lifetimes of BPs with repetitive brightenings, e.g., the persistent flashers. If a BP reappears within $30$ pixels (i.e., about twice our BP's size threshold) of its original location within $160~sec$ we consider it to be the same BP. If these conditions are not met then we assume the new brightening to belong to a new BP.\\
The relatively short time series biases the measurements towards shorter lifetimes and hence the lifetime distribution must be corrected (see Sect.~\ref{subsec-lifetimeHist}).
Two possible sources of uncertainty of the measured lifetime are the exposure time of the SuFI Ca~{\sc ii}~H images (i.e., $1~sec$) and the observing cadence. The latter parameter differs for different time series used in this study and thus we consider an average value of the cadences presented in Table~\ref{table:obslog}, weighted by the number of BPs identified in each image sequence. Therefore, the Ca~{\sc ii}~H BPs' lifetimes reported in this paper have an overall uncertainty of $9~sec$.

\subsubsection{Intensity}

The mean value of the intensity inside the area of each detected BP is referred to as its absolute intensity. The intensity values are measured from the original phase-diversity reconstructed Ca~{\sc ii}~H images, by referencing their accurate positions and sizes obtained from the restored images (described as in Sect.~\ref{subsec-restoration}, Sect.~\ref{subsubsec-location} and Sect.~\ref{subsubsec-size}). The contrast of the BP is then determined by normalising its intensity to the mean intensity of the entire Ca~{\sc ii}~H frame that the BP is observed in, i.e., we form the contrast relative to the average quiet-Sun.

\subsubsection{Horizontal Velocity}
\label{subsec-Hvelocity}

The instantaneous horizontal velocity ($v$) was determined by taking the difference between the locations of a BP (known to sub-pixel accuracy) in two consecutive frames and dividing by the time between frames. The statistical uncertainty in measuring this frame-to-frame horizontal velocity ($\sigma_{v}$) was computed from the error propagation analysis.\\
We can then assess the mean value of the horizontal velocity $\langle v \rangle$ and its uncertainty for each BP, consisting of $n$ time steps (frames), using the weighted mean~\citep{Wall2003},

\begin{equation}
\left \langle v \right \rangle=\frac{\sum_{j=1}^{n}\left ( v_{j}/ {\sigma _{j}}^{2} \right )}{\sum_{j=1}^{n}\left ( 1/ {\sigma _{j}}^{2} \right )}\,,
	\label{equ:meanV}
\end{equation}
\noindent
where $v_{j}$ and ${\sigma_{j}}^2$ are the $j$th determined instantaneous horizontal velocity ($v$) and its variance, respectively. The best estimate of the uncertainty of $\langle v \rangle$ is then,

\begin{equation}
\sigma _{\left \langle v \right \rangle}=\sqrt{\frac{1}{\sum_{j=1}^{n}\left ( 1/ {\sigma_{j}}^{2} \right )}}\,;
	\label{equ:sigmaV}
\end{equation}
\noindent
We also use Eqs.~(\ref{equ:meanV}) and (\ref{equ:sigmaV}) to calculate the total mean value of the horizontal velocity and its uncertainty. Note that we do not take the normal average of the velocity values to determine the mean velocities (for each BP or the mean over all BPs), since different velocity values have different uncertainties. This latter effect arises because of different distances a BP may travel in equal time steps and/or different time steps due to ($1$) a variable cadences within an image sequence caused by missing or discarded images, and/or due to ($2$) a temporarily undetected BP. Such an analysis results in an average uncertainty (weighted average over all BPs) of $0.02~km~s^{-1}$ in the horizontal velocities.

\subsection{Magnetic Field}

The magnetic properties of the Ca~{\sc ii}~H BPs were investigated by comparing with maps of $CP$ and $LP$ determined as described in Sect.~\ref{subsec-stokes}. In order to avoid noise introducing false detections, isolated single pixels lying above $3\sigma$ were not counted as signal.
The magnetic regions lying above $3\sigma$ are outlined in Fig.~\ref{fig:bp_temp} by red and yellow contours for opposite polarities of the $CP$. 
A Ca~{\sc ii}~H BP is then considered magnetic if it overlaps with the computed photospheric magnetic patches for more than $50\%$ of its lifetime. The magnetic field is generally still present at times when it is not associated with the BP but either lies below the $3\sigma$ threshold, or just fails to overlap with the BP. The BP shown in Fig.~\ref{fig:bp_temp} seems to be almost completely inside a magnetic patch of negative polarity for most of the time. However, there are times (e.g., at the time $3041~sec$ from UT midnight) when this BP is observed just outside the boundary of the magnetic patch.\\
The peak values of $CP$ and $LP$ over a BP's area is considered as corresponding polarisation degrees of the BP measured from the spatially smoothed non-reconstructed data (obtained as described in Sect.~\ref{subsec-stokes}).

\subsection{Inversions}
\label{subsec-inversion}
We use the results of SIR code (Stokes Inversion based on Response functions;~\citealt{Ruiz-Cobo1992}) inversions of the IMaX Fe I $5250.2~\AA$ Stokes vector carried out by L.~R.~Bellot~Rubio. The code numerically solves the radiative transfer equation (RTE) along the line of sight (LOS) for the Zeeman-induced polarisation of light under the assumption of local thermodynamic equilibrium (LTE). The code deals with all four Stokes parameters ($I$, $Q$, $U$, $V$), any combinations of which are fitted for any arbitrary number of spectral lines. Then, the differences between the observed and synthetic Stokes profiles are iteratively minimized by modifying an initial model atmosphere.\\
Details of the employed SIR inversions can be found in the paper by~\citet{Guglielmino2011}.\\
The Harvard Smithsonian Reference Atmosphere (HSRA;~\citealt{Gingerich1971}) is used as the initial model atmosphere. The temperature has two nodes and the inversions recover the temperature stratification in layers between $log(\tau)~-4$ and $0$ (where $\tau$ is the optical depth of the continuum at $5000~\AA$). Stray light is not taken into account for this inversion and a magnetic filling factor of equal to unity is assumed. This implies that the magnetic field strength from the SIR code is a lower limit of the true value.

\section{Results and Discussion}

A total of $107$ Ca~{\sc ii}~H BPs were identified in $6$ different image sequences obtained by the SuFI instrument, where each BP is counted only once during its lifetime. We found a relatively low BP average number density of $0.03~(Mm)^{-2}$ corresponding to $\approx1.8\times10^5$ BPs at any given time on the whole solar surface. This number is roughly an order of magnitude smaller than the number of granules on the solar surface at any given time. This low number can be due to the rather restrictive identification criteria, in particular that we consider only small isolated BPs (comparable in size with the {\sc Sunrise} spatial resolution) that do not merge or split while they migrate. These criteria restrict our sample mainly to internetwork BPs. The parameters intensity contrast, horizontal velocity and size, were calculated for each BP at every time step. The lifetime was measured only for $47$ BPs whose birth and death times were both observable. The BP is considered magnetic if the average of the local Stokes $V$ signal within a $3\times3$ pixel box around its corresponding location is higher than the $3\sigma$ level over at least $50\%$ of the detected BP lifetime.

\subsection{A Case Study: Tracking an individual Ca~{\sc ii}~H BP}
\label{subsec-caseStudy}

\begin{figure}[h!]
	\centering
	\includegraphics[width=8.5cm]{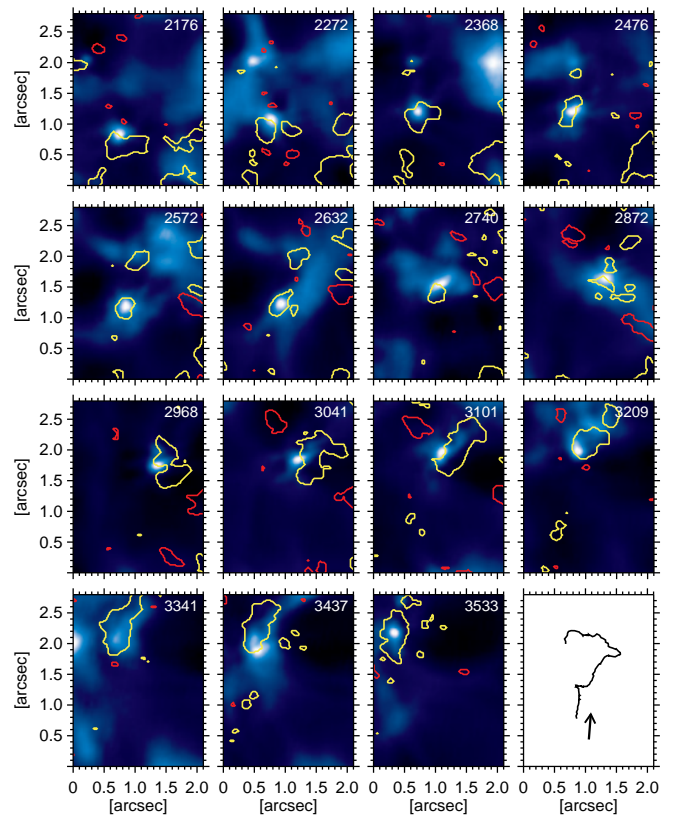}
	\caption{Temporal evolution of a selected BP (Lifetime: $1321~sec$), observed on 9 June 2009 in the {\sc Sunrise}/SuFI Ca~{\sc ii}~H passband. Time proceeds row by row and from left to right within each row. The red and yellow contours outline magnetic regions with opposite polarities of the line-of-sight magnetic field component and match the $3\sigma$ level in Stokes $V$. In the bottom-right panel, the trajectory of the BP is plotted. The arrow marks the sense of the motion. The time from UT midnight, in seconds, is shown in the upper-right corner of each frame. Note the irregular intervals between frames.
              }
	\label{fig:bp_temp}
\end{figure}

\begin{figure}[htbp!]
	\centering
	\includegraphics[width=8cm]{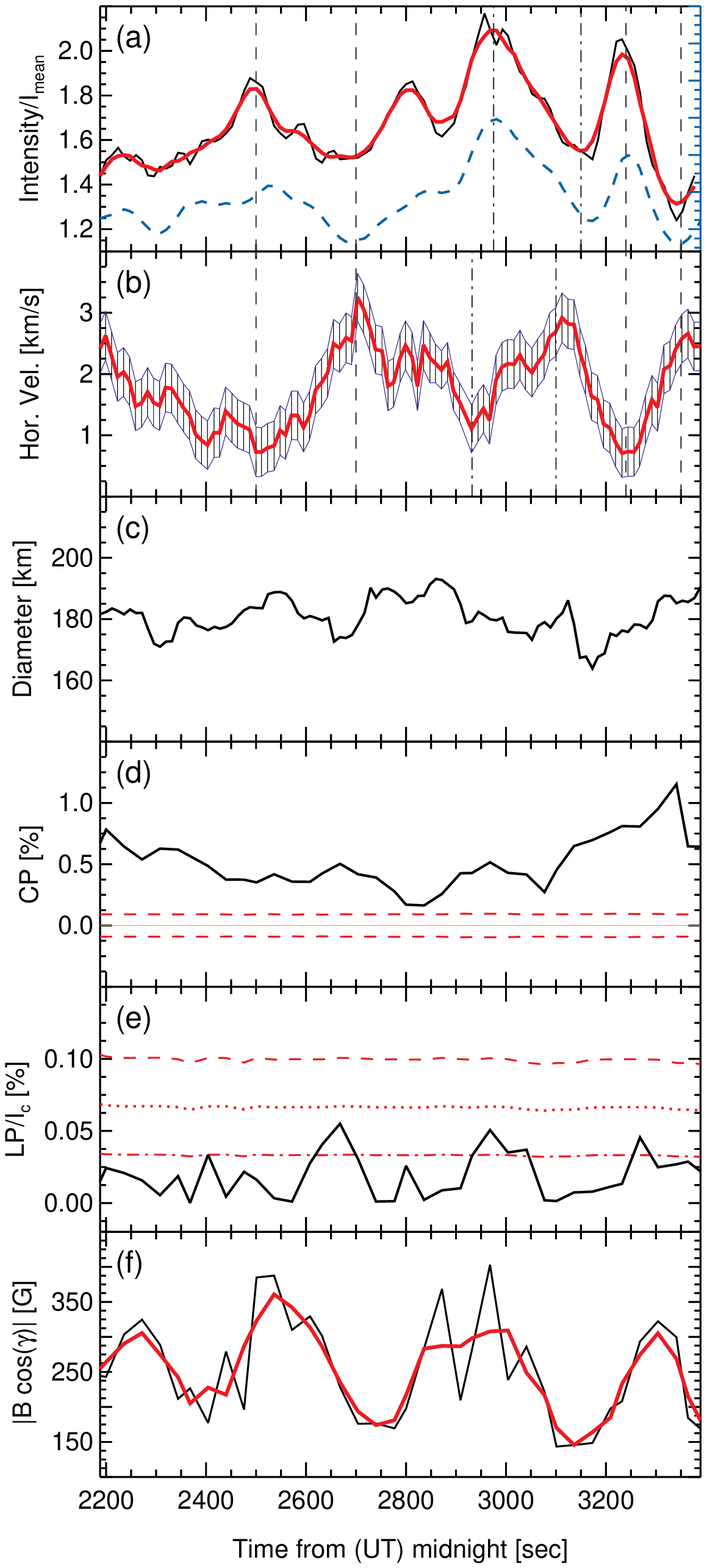}
	\caption{Evolution of ($a$) the intensity contrast, ($b$) horizontal velocity, ($c$) diameter, ($d$) $CP$, a measure of the strength of Stokes $V$, ($e$) total net linear polarisation $LP$, and ($f$) field strength of the selected BP shown in Figs.~\ref{fig:obs_examples} and \ref{fig:bp_temp}. $CP$ and $LP$ are normalised to Stokes $I$ continuum. In panel ($a$) and ($f$), the thin curves respectively represent the time variation of average intensity and the unsigned longitudinal magnetic field within the BP while the smoothed thick lines (red in online version) show a 5-frames boxcar average. They are both normalised to the mean intensity of the image. The (blue) dashed curve in panel ($a$) represents the smoothed average intensity within the BP, in arbitrary unit, from the restored images (see main text). Evolution of horizontal velocity, smoothed within 5 frames by weighted averaging, is plotted as the thick line (red in the online version) in panel ($b$) while the thin vertical solid lines indicate the error bars. The vertical dashed lines in the two upper panels serve to guide the eye to the general anti-correlation between intensity and horizontal velocity (see main text). The $3\sigma$ noise levels for both $CP$ and $LP$, are drawn as dashed-lines in panels ($d$) and ($e$), while the dotted and dot-dashed lines in panel ($e$) are the $2\sigma$ and $1\sigma$ noise levels, respectively. The unsigned longitudinal magnetic field depicted in panel ($f$) was computed from the SIR inversions (see main text).
              }
	\label{fig:bp_var}
\end{figure}

Here, we study the evolution of one of the identified BPs in detail. Snapshots of the Ca~{\sc ii}~H intensity in the region around the BP as well as overlaid contours of $CP$ are plotted in Fig.~\ref{fig:bp_temp}. The BP moves within an area of roughly $1 \times 2~arcsec^2$ in the course of its lifetime of $1321\pm13~sec$ (see its trajectory in the bottom right frame). The BP is located almost always inside a magnetic patch, except for a few moments such as the one that occurs at time $3041~sec$ (all times are given relative to UT midnight). At this time, the BP lies mainly outside the boundary of the magnetic patch, however, a small part of its area overlaps with $CP>3$ signal. Note the irregular intervals between the frames. Fig.~\ref{fig:bp_var} illustrates the temporal behaviour of physical parameters of the BP. It has a mean intensity of $1.65\pm0.01~\langle I_{Ca} \rangle$ and moves with a mean proper motion velocity of $1.86\pm0.08~km~s^{-1}$.

\noindent
The frame-by-frame intensity variation (thin curve) is plotted in panel $(a)$. The intensity curve smoothed by a boxcar averaging of $5$ frames is overplotted (red in the online version). In order to estimate the effect of transient brightenings, e.g., passing waves, on the BP's intensity variation, we overplot the variation of the averaged intensity within the BP (smoothed by a 5-frames boxcar averaging; blue dashed curve) measured in the restored images, i.e., the images from which the larger wave-like features are subtracted (see Sect.\ref{subsec-restoration}); the curve is in arbitrary units. A comparison between the two smoothed intensity curves validates the intrinsic intensity variation of the BP. Panel $(b)$ represents the variation of horizontal velocity smoothed over $5$ frames by weighted averaging, described in Sect.~\ref{subsec-Hvelocity} (thick curve; red in the online version). The thin vertical solid lines indicate the error bars. The values of intensity and horizontal velocity vary by a factor of nearly $2$ and $3$, respectively.\\
Close inspection of panels $(a)$ and $(b)$ reveals an anti-correlation between these two quantities for the BP. The vertical dashed lines mark the times when the intensity maxima and minima fall together with minima and maxima in horizontal velocity, respectively. However, two peaks in the horizontal motion plot and the corresponding dips in intensity plot, indicated by vertical dot-dashed lines, show a small time delay. An anti-correlation between the intensity and horizontal velocity is also observed in some other BPs, sometimes also with a small time delay. All BPs do not show it.\\
Panel $(c)$ in Fig.~\ref{fig:bp_var} illustrates the evolution of the BP's diameter, which is found to be constant to within $10-15~\%$.

\begin{figure}[h]
    \centering
	\includegraphics[width=8.5cm, trim = 0 0 0 0, clip]{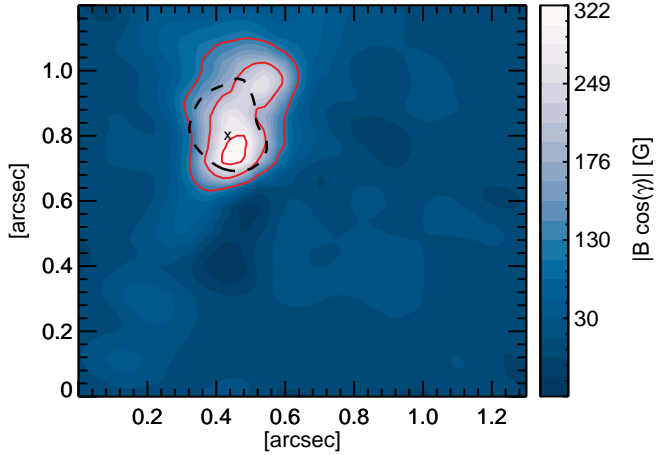}
	\caption{Map of the unsigned longitudinal component of magnetic field, $\vert B \cos(\gamma)\vert$, centred on the BP shown in Fig.~\ref{fig:bp_temp} at time $3304~sec$ from UT midnight. The solid contours indicate the height-independent $\vert B \cos(\gamma)\vert$ at $100, 200$ and $300~ G$. The Ca~{\sc ii}~H BP 's area is indicated by the dashed contour.}
	\label{fig:bp_Bmap}
\end{figure}

\noindent
The $CP$ (obtained as described in Sect.~\ref{subsec-stokes} and normalized to the continuum intensity $\langle I_{c} \rangle$ averaged over the full IMaX FOV), plotted in panel $(d)$, is always above the $3\sigma$ noise level, which demonstrates the magnetic origin of this BP.\\
The variation of total linear polarisation ($LP$; see Sect.~\ref{subsec-stokes}), normalised to $\langle I_{c} \rangle$, is plotted in panel $(e)$. The dashed and dotted lines in panel $(e)$ show the $3\sigma$ and $2\sigma$ noise levels. The $LP$ is almost always below the $1\sigma$ level (indicated by the dot-dashed line), implying that the linear polarisation signal detected at the position of the small BP is not significant at the $1\sigma$ level. Nevertheless, inversions return a strongly inclined field for such a BP with inclination angles between $78$ and $90^{\circ}$ in the course of its lifetime. This result, however, is probably affected by noise as we find by comparing with inclinations derived from the locations of the BP in images sampling different heights in the atmosphere (Jafarzadeh et al., in preparation).\\
The values of the field strength multiplied by the absolute value of the cosine of the inclination angle $\vert B \cos(\gamma)\vert$ (i.e., the unsigned longitudinal component of the magnetic field) returned by the SIR inversion code are plotted in panel ($f$). We plot the variations of $\vert B \cos(\gamma)\vert$ since this value is retrieved more reliably than B (see Sect.~\ref{subsec-BcosGamma}). $\vert B \cos(\gamma)\vert$ varies between $108$ and $361~G$ assuming that the magnetic feature is spatially resolved. If this particular BP is spatially unresolved by {\sc Sunrise}~\citep{Riethmuller2012}, which is likely given the low $CP$ value, these field strength values are the lower limits. Fig.~\ref{fig:bp_Bmap} shows a $\vert B \cos(\gamma)\vert$ map of the BP at time $3304~sec$ from UT midnight. $\vert B \cos(\gamma)\vert$ at the position of this small internetwork BP reaches a maximum value of $322~G$ at this point in time. The over-plotted solid contours show the field strength levels of $100, 200$ and $300~ G$. The location of the Ca~{\sc ii}~H BP is overlaid (dashed contour). The maximum intensity of the the BP, indicated by a cross, is offset by $0.05~arcsec$ ($20\%$ of the diameter of the BP) relative to the location of the highest $\vert B \cos(\gamma)\vert$. This offset may represent a small inclination of the flux tube from the vertical direction, since the Ca~{\sc ii}~H brightness and $\vert B \cos(\gamma)\vert$ are determined at different heights. We will consider inclinations of magnetic elements in a separate paper. A small observing time difference between the SuFI Ca~{\sc ii}~H and IMaX images can also be a reason for such an offset.

\subsection{Persistent Flasher}
\label{subsec-flasher}
\begin{figure}[h]
    \centering
	\includegraphics[width=7.5cm]{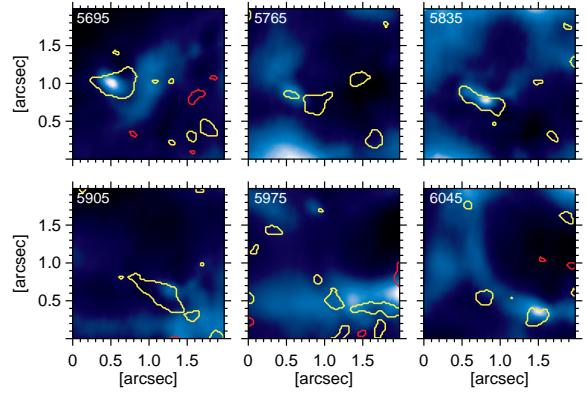}
	\caption{Temporal evolution of a BP similar to the persistent flasher of~\citet{Brandt1992}. Time proceeds row by row from left to right in each row. The yellow and red contours indicate magnetic patches of opposite polarities. The time in seconds, from UT midnight, is given in the upper-left corner of each frame.}
	\label{fig:flasher}
\end{figure}

\begin{figure}[h!]
	\centering
	\includegraphics[width=8cm]{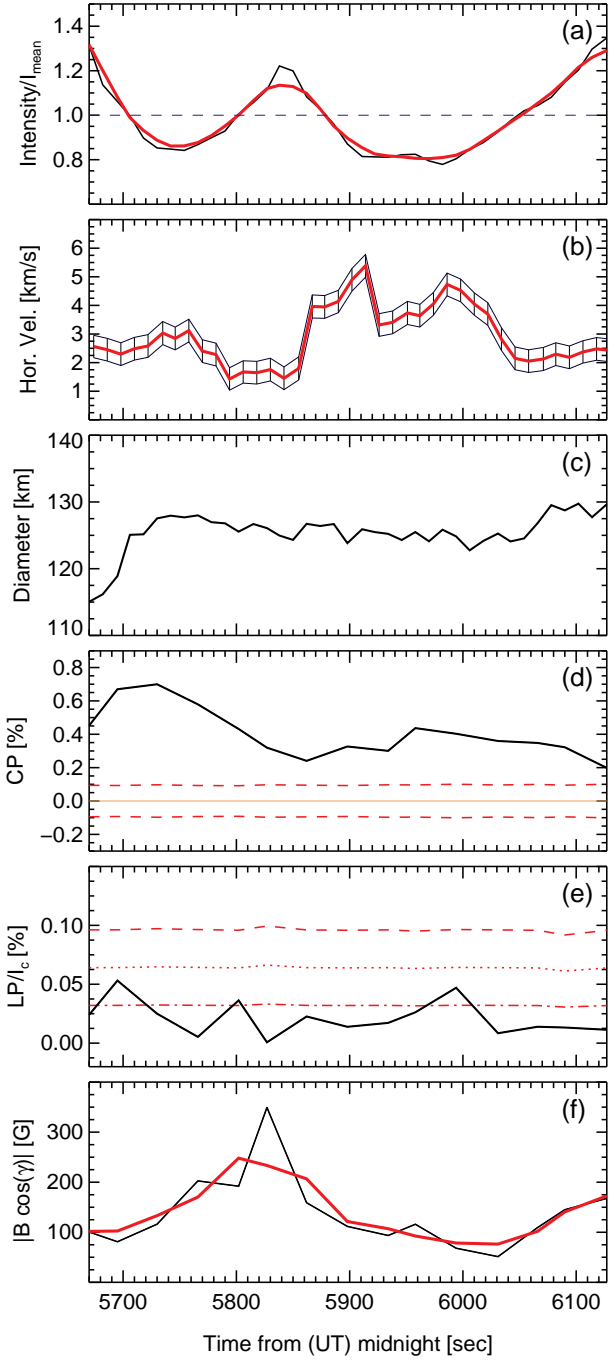}
	\caption{The same as Fig.~\ref{fig:bp_var}, but for the persistent flasher shown in Fig.~\ref{fig:flasher}.
              }
	\label{fig:flasher_var}
\end{figure}

\noindent
Among the $107$ analysed BPs $7$ are similar to those named persistent flashers by~\citet{Brandt1992}. Their main difference to the majority of the selected BPs is the occurrence of intensity variations about the detection limit ($\langle I_{Ca} \rangle$) which give rise to the apparent succession of flashes.\\
An example of such a BP is shown in Fig.~\ref{fig:flasher}, which shows the Ca~{\sc ii}~H intensity at $6$ times within an area of $2\times2~arcsec^2$. The contours outline magnetic regions with $CP$ above the $3\sigma$ level. The temporal variations of different parameters of the BP are plotted in Fig.~\ref{fig:flasher_var}. From panel $(a)$ it is evident that between two bright phases the intensity of the BP drops below the intensity limit of the detection giving it the appearance of flashing, although its intensity variations are no larger, in a relative sense, than those of the BP displayed in Fig~\ref{fig:bp_var}. Interestingly, an anti-correlation between the intensity and the horizontal velocity in panel $(b)$ is also seen for this persistent flasher. The flasher travels with a mean proper motion velocity of $2.82\pm0.06~km~s^{-1}$. The size is constant within a range of $10$\% (panel $(c)$). We cannot rule out that the large excursion of the size at the beginning of the plotted time series is due to a false detection at the very beginning of the BP's identification. Similar to the other BPs under study, $CP$ always lies above the $3\sigma$ noise level, also for the persistent flasher; see Fig.~\ref{fig:flasher_var}$d$. Again, the LP signal lies almost always below the $\sigma$ level (dot-dashed line). Therefore, the same conclusion as for the BP studied in Sect.~\ref{subsec-caseStudy} can be made here of the field inclination. The $\vert B \cos(\gamma)\vert$, plotted in panel $(f)$, has a mean value of $190~G$, smaller than the majority of the normal, non-flasher BPs under study.

\noindent
In this investigation, the few ``persistent flashers'' ($6$\% of all identified BPs) clearly have a magnetic origin, with horizontal velocities of $2.4~km~s^{-1}$ on average. This makes the persistent flashers about equally mobile as the other Ca~{\sc ii}~H BPs (see Sect.~\ref{subsubsec-hvel}). The flashers differ from the normal BPs mainly in that they (the flashers) tend to be less bright. The persistent flashers are on average $47\%$ less bright than the other BPs studied here (see Sect.~\ref{subsubsec-int}). This causes the occurrence of intensity variations ($\approx$ of a factor of $3$) around our detection limit and hence the intermittent appearance of the BP. We conclude that persistent flashers are less bright, but otherwise normal magnetic BPs.

\subsection{Statistical Studies}
\label{subsec-statistics}

Table~\ref{table:BPnum} summarises the number and mean values of the BPs studied here. The mean lifetime value is to our knowledge the longest obtained for Ca~{\sc ii}~H BPs so far (see Sect.~\ref{subsec-lifetimeHist}), although it is a lower limit due to the finite lengths of the employed time series. This reflects the excellent quality of the {\sc Sunrise}/SuFI observations. The mean diameter is influenced by our criteria to only concentrate on the isolated and point-like features which do not display any internal fine structure.

\begin{table}[hb]
\caption{Summary of averaged properties of the Ca~{\sc ii}~H BPs}     
\label{table:BPnum}      
\centering                   
\begin{tabular}{l c c}      
\hline\hline              
Parameter & Mean value$~^{\mathrm{ a}}$ & No. of BPs\\   
\hline                      
   Intensity & $1.48\pm0.3~\langle I_{Ca} \rangle$ & $107$\\   
   Horizontal Velocity & $2.2\pm1.8~km~s^{-1}$ & $107$\\
   Diameter & $0.2\pm0.02~arcsec$ & $107$\\
   Lifetime & $673\pm526~sec$ & $47$\\
   $CP~^{\mathrm{ b}}$ & $0.32\pm0.21~\%$ & $53$\\
   $\vert B \cos(\gamma)\vert~^{\mathrm{ b}}$ & $142\pm87~G$ & $53$\\
\hline                           
\end{tabular}
\begin{list}{}{}
\item[$^{\mathrm{a}}$] The uncertainty estimates were calculated from the standard deviations of the parameters' distributions. For formal errors see main text.
\item[$^{\mathrm{b}}$] From {\sc Sunrise}/IMaX\\
$CP$: Circular polarisation degree, described as in Sect.~\ref{subsec-stokes}; $\vert B \cos(\gamma)\vert$: unsigned longitudinal component of magnetic field.

\end{list}
\end{table}

\begin{figure}[h!]
	\centering
	\includegraphics[width=8.5cm]{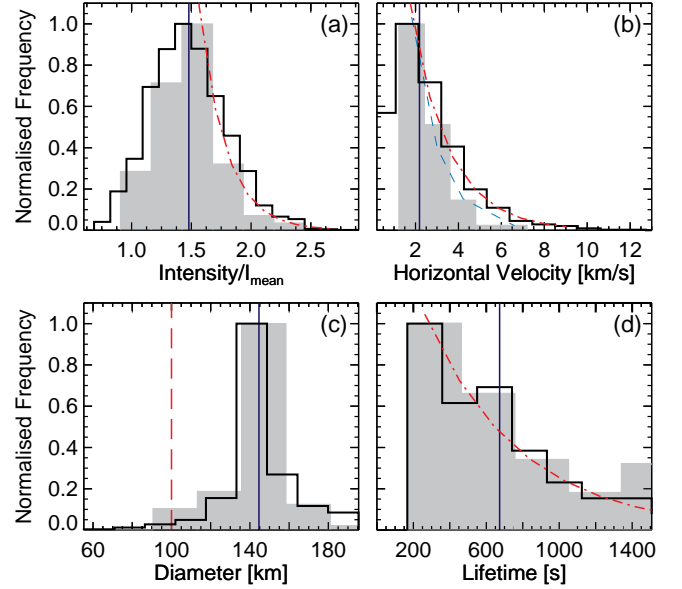}
	\caption{Statistical properties of the detected {\sc Sunrise} Ca~{\sc ii}~H BPs.
	($a$): Distribution of mean intensity contrast for all 5677 individual measurements (black, outlined histogram) and of the time averaged contrast of all 107 BPs (grey, filled histogram). 
	($b$): Distribution of all 5570 individual measurements of horizontal velocity (black, outlined histogram) and of the values averaged over their respective lifetimes of the 107 studied BPs (grey, filled histogram).
	($c$): Histogram of mean size (diameter) of all 107 BPs (grey, filled histogram) and all 5677 individual measurements (black, outlined histogram). The vertical dashed line indicates the spatial resolution achieved by  the {\sc Sunrise} telescope.
	($d$): Histogram of lifetimes of 47 BPs (black, outlined histogram) and of corrected lifetimes (grey, filled histogram; see main text).
	The vertical solid lines indicates the mean values of the distributions.
	The fits (dot-dashed and dashed curves) are explained in the main text.}
	\label{fig:stats}
\end{figure}

\noindent
Distributions of the intensity, horizontal velocity and size of all $107$ BPs, as well as the lifetime of the $47$ BPs for which it is available, are presented in Fig.~\ref{fig:stats}. Before discussing these properties in detail, we point out that all of the BPs for which simultaneous polarimetric observations were available ($\approx 50\%$ of all BPs) are clearly associated with magnetic fields (see Sect.~\ref{subsec-CP}). It is useful to bear this in mind when reading the following sections.\\
The difference in the spatial resolutions and scattered light between different sets of data in the literature is considerable and may well be a dominant cause of discrepancies between various studies. In addition, observing in different wavelengths and hence sampling different heights in the solar atmosphere affects the results. The same is true for the variation in the lengths of observed image sequences. Finally, differences in definition of BPs may also influence the results (e.g., thresholds used).\\
Mean values of BP physical parameters determined here are compared with past findings in Table~\ref{table:comp}. In this table, because of the common origin and the similarity in appearance of the magnetic BPs in $G$-band, TiO-band and Ca~{\sc ii}~H/K passbands~\citep{de-Wijn2008}, we have also included $G$-band and TiO passbands, which normally sample lower regions of the solar atmosphere than Ca~{\sc ii}~H/K. However, the sampled heights depend on the filter widths. Thus, the measured Ca~{\sc ii}~H/K signal also gets contributions from the lower layers of the atmosphere. The average horizontal speed of non-magnetic K-grains is also listed in this table in order to illustrate the difference to magnetic BPs.\\

\subsubsection{Intensity}
\label{subsubsec-int}

The distribution of all $5677$ measured instantaneous intensity contrasts of the $107$ BPs is represented by the outlined histogram in Fig.~\ref{fig:stats}$a$. Its exponential fit with an e-folding width of $0.22~\langle I_{Ca} \rangle$ is overlaid as a red, dot-dashed line (fit limited to the range of $1.45$ to $2.7~\langle I_{Ca} \rangle$).\\
The intensity contrast distribution shows a lower limit of $0.69~\langle I_{Ca} \rangle$ and the brightest detected BP reaches a maximum value of $2.86~\langle I_{Ca} \rangle$. The mean contrast value, averaged over all BPs and all time steps is $1.48~\langle I_{Ca} \rangle$. The few intensity values less than $\langle I_{Ca} \rangle$ belong to persistent flashers whose intensity drops below the detection limit ($\langle I_{Ca} \rangle$) at some moments of time, so that the average is low, although in between they are clearly visible.\\
The filled histogram in Fig.~\ref{fig:stats}$a$ is the distribution of the lifetime average of intensity contrast of each individual BP of all $107$ BPs.\\
A comparison with values in the literature is summarised in Table~\ref{table:comp}. Such a comparison is not straightforward because observations at different wavelengths, with different filter widths and different spatial resolution give a range of contrasts. Also, none of the non-{\sc Sunrise} studies of Ca~{\sc ii}~H or K BPs provides their average contrasts.\\
~\citet{Berger1995} showed that the intensity of their GBPs, on average, reaches $1.27$ relative to the quiet-Sun average. A range of $0.8-1.8~\langle I_{QS} \rangle$ was reported for the same wavelength by~\citet{Sanchez-Almeida2004} at an estimated spatial resolution of $135~km$ on the Sun. Note that the lower boundary of this range means that these BPs are darker than $\langle I_{QS} \rangle$ implying that the GBPs were bright only with respect to the intergranular lanes surrounding them. A mean value $1.17~\langle I_{QS} \rangle$ was obtained by~\citet{Mostl2006} for magnetic GBPs.\\
\citet{Riethmuller2010} studied the peak contrast of the BPs observed in different passbands of the {\sc Sunrise} observatory between $2140~\AA$ and $5250.2~\AA$, including Ca~{\sc ii}~H $3968~\AA$. They found mean values of the peak contrast equal to $1.11$, $1.35$, $1.52$, $1.60$, $1.89$ and $2.31~\langle I_{QS} \rangle$ for 5250, 3120, 3000, 3880, 3968 and $2140~\AA$, respectively. Their mean contrast value of the Ca~{\sc ii}~H BPs is somewhat larger than the $1.48$ we measured in this study. This difference partly reflects the different criteria employed to identify the BP studied (we restrict ourselves to smaller BPs). Moreover, they measured the peak contrast of each BP, which is expected to be higher than the contrast averaged over a BP's area as we determined.\\

\begin{table*}[ht]
\caption{Comparison of the mean values of properties of BPs obtained in this study with values in the literature.}
\label{table:comp}     
\centering                          
\begin{tabular}{l c c c c c c c c}       
\hline\hline                 
Reference & Telescope & Passband & Pol. & Feature$^{\mathrm{ b}}$ & Intensity & H. Velocity & Diameter & Lifetime\\    
 & (Diffraction & (FWHM) & Info.$^{\mathrm{ a}}$ &  & [$\langle I_{QS} \rangle$] & [$km~s^{-1}$] & [$arcsec$] & [$s$]\\
  & Limit) &  &  &  &  &  &  & \\   
\hline                        
This study & {\sc Sunrise} (0\farcs10) & Ca~{\sc ii}~H (1.8 \AA) & Yes & IBP & $1.48$ & $2.2$ & $0.2$ & $673$\\      
\citet{Riethmuller2012} & {\sc Sunrise} (0\farcs10) & CN (8.0 \AA)$^{\mathrm{ c}}$ & Yes & MBP & - & - & 0.45 & -\\
\citet{Keys2011} & DST (0\farcs12) & $G$-band (9.2 \AA) & No & MBP & - & 1.0 & - & 91\\
\citet{Abramenko2010} & NST (0\farcs11) & TiO (10 \AA) & No & IBP & - & - & $\approx$0.16 & $<120$\\
\citet{Crockett2010} & DST (0\farcs12) & $G$-band (9.2 \AA) & No & MBP & - & - & 0.31 & -\\
\citet{Riethmuller2010} & {\sc Sunrise} (0\farcs10) & Ca~{\sc ii}~H (1.8 \AA)$^{\mathrm{ c}}$ & Yes & MBP & $1.89$ & - & - & -\\
\citet{Mostl2006} & SST (0\farcs1) & $G$-band (11.6 \AA) & No & MBP & 1.17 & 1.11 & 0.28 & 263\\
\citet{de-Wijn2005} & DOT (0\farcs2) & Ca~{\sc ii}~H (1.3 \AA) & No & IBP & - & - & - & 258\\
\citet{Sanchez-Almeida2004} & SST (0\farcs1) & $G$-band (10.8 \AA) & No & GBP & 0.8-1.8 & - & 0.25 & 225\\
\citet{Nisenson2003} & DOT (0\farcs2) & $G$-band (10 \AA) & No & GBP & - & 0.89$^{\mathrm{ d}}$ & - & 552\\
\citet{Berger2001} & SVST (0\farcs2) & $G$-band (12 \AA) & Yes & GBP & - & 1-5$^{\mathrm{ e}}$ & 0.145 & -\\
\citet{Wellstein1998} & VTT (0\farcs3) & Ca II K (0.3 \AA) & No & NBP & - & 6.6 & - & -\\
\citet{Steffens1996} & VTT (0\farcs3) & Ca II K (0.3 \AA) & No & K-grain & - & 50 & - & -\\
\citet{Berger1995} & SVST (0\farcs2) & $G$-band (12 \AA) & No & GBP & 1.27 & - & 0.35 & -\\
\citet{Muller1994} & Pic du Midi (0\farcs25) & 5750 \AA ~(60 \AA) & No & NBP & - & 1.4 & - & 1080$^{\mathrm{ f}}$\\
\citet{Soltau1993} & VTT (0\farcs3) & Ca II K (0.3 \AA) & No & IBP & - & 2.5 & - & -\\
\citet{Muller1983} & Pic du Midi (0\farcs25) & 5750 \AA ~(60 \AA)$^{\mathrm{ c}}$ & No & IBP & - & - & - & 540\\
\hline                                   
\end{tabular}
\begin{list}{}{}
\item[$^{\mathrm{a}}$] Polarisation (magnetic field) information
\item[$^{\mathrm{b}}$] IBP: Internetwork Bright Point; NBP: Network BP; MBP: Magnetic BP; GBP: $G$-band BP
\item[$^{\mathrm{c}}$] See main text for other passbands
\item[$^{\mathrm{d}}$] rms velocity
\item[$^{\mathrm{e}}$] From~\citet{Berger1996}
\item[$^{\mathrm{f}}$] From~\citet{Muller1983}, who found this value for NBPs.
\end{list}
\end{table*}

\subsubsection{Horizontal Velocity}
\label{subsubsec-hvel}

The distribution of the instantaneous proper motion velocity of all individual $5570$ frame-to-frame measurements made in this study is given by the outlined histogram in Fig.~\ref{fig:stats}$b$. The red, dot-dashed line is the exponential fit to this histogram (for velocities larger than $1.8~km~s^{-1}$) with an e-folding width corresponding to a horizontal velocity of $1.8~km~s^{-1}$. However, lower velocities ($<1~km~s^{-1}$) are under-represented relative to the exponential fit, suggesting that the BPs tend not to stay at a given spot. The distribution displays a tail reaching up to $10~km~s^{-1}$. There are a few measurements at higher velocities which lie outside the range of this histogram reaching upto $15.5~km~s^{-1}$.
Using the values of temperature, density and gas pressure of the VAL-C atmospheric model~\citep{Vernazza1981} at the average formation height of the {\sc Sunrise} Ca~{\sc ii}~H BPs ($500~km$), the sound speed is estimated to be between $6.5$ and $7~km~s^{-1}$. Thus, Fig.~\ref{fig:stats}$b$ implies that about $3.5\%$ of the time the BPs' motion is supersonic.\\
The distribution of the mean value (lifetime average) of the horizontal velocity of each BP is over-plotted (filled histogram). The exponential fit to this histogram (with an e-folding width corresponding to a horizontal velocity of $1.3~km~s^{-1}$) is plotted as a blue, dashed line. Power law fits were found to be unacceptable for both these distributions. The statistics reveal that the mean horizontal velocity of the BPs studied in this work varies between $1.2$ and $6.6~km~s^{-1}$ (mean values for individual BPs) with an average value of $2.2~km~s^{-1}$ (averaged over all BPs).\\
We remind the reader that the value of the frame-to-frame velocity is biased by the measurement error in positioning the BPs (see Sect.~\ref{subsubsec-location}). An upper limit on the velocity induced by this error is about $0.7~km~s^{-1}$. The effect of this error is greatly decreased when computing the velocities averaged over the lifetimes of individual BPs, as mentioned above (see Sect.~\ref{subsec-Hvelocity}).\\
As summarised in Table~\ref{table:comp}, the measured horizontal velocities of the Ca~{\sc ii}~H BPs are larger than the velocities of magnetic BPs in the photosphere, which are quoted as $0.06~km~s^{-1}$~\citep{vanBallegooijen1998}, $1-1.4~km~s^{-1}$ ~\citep{Mostl2006}, although higher horizontal velocities of up to $5~km~s^{-1}$ for a few indicated BPs have been reported in the literature~\citep{Berger1996}. Note that \citet{vanBallegooijen1998} determined horizontal velocities by tracking corks so that these refer to averaged motions of clusters of BPs instead of isolated ones. However, our value is significantly smaller than the average of $6.6~km~s^{-1}$~\citep{Wellstein1998} and $49~km~s^{-1}$~\citep{Steffens1996} reported for the Ca~{\sc ii}~H internetwork K-grains as well as $7-10~km~s^{-1}$ horizontal velocities of the network BPs found by~\citet{Wellstein1998}. Whereas the somewhat larger values we find compared to the GBPs may be acceptable (e.g., caused by swaying of magnetic flux tubes), the huge range of values quoted for the chromosphere (Ca~{\sc ii}~H and K observations) shows how difficult it is to compare these values. Obviously, the different measurements refer to very different kinds of features. In particular, K-grains refer to a wave phenomenon and cannot be compared with the magnetic features studied here. There still remains the discrepancy with the large value for network BPs published by~\citet{Wellstein1998}. Possible explanations for this are: ($a$) Network BPs (NBPs) move much faster than internetwork BPs (IBPs). ($b$) Because \citet{Wellstein1998} had no magnetic information, their measurements may have been affected by misidentification of passing waves even in the case of NBPs. We obtained stable results only after applying a careful filtering technique. ($c$) Proper motion velocities derived from ground-based observations tend to be exaggerated due to differential seeing-induced deformations in different frames of the time series.\\
Since we are unaware of any studies (e.g., in the $G$-band) suggesting higher horizontal velocities of network BPs than internetwork BPs, and due to the much larger horizontal velocities found by \citet{Wellstein1998} than of photospheric BPs, we conclude that explanations ($b$) and ($c$) are more likely than ($a$).

\subsubsection{Size}
\label{subsec-sizeHist}

The distribution of the BP diameters obtained from all $5677$ individual measurements are represented by the outlined histogram in Fig.~\ref{fig:stats}$c$. The BPs' diameters range between $85$ and $195~km$ or $\approx0.11$ and $\approx0.27~arcsec$, with a mean value of $150\pm15~km$ ($\approx0.2\pm0.02~arcsec$).\\
The filled gray histogram represents lifetime averages. Both histograms exhibit a rather narrow peak centred at $\approx150~km$ corresponding to a diameter of $\approx0.2~arcsec$. This narrowness of the peak reflects our size criterion in which a {\sc Sunrise} Ca~{\sc ii}~H BP was defined as a bright, point-like feature with a maximum diameter of $0.3~arcsec$ at any given time. On the other hand, the BPs' size distribution extends down to the resolution reached by {\sc Sunrise} of roughly $100~km$ in the best time series. Therefore, the typical diameter of the BPs studied here is between $0.15$ and $0.25~arcsec$. Possible explanations for the missing smaller BPs in our statistics (down to the theoretical resolution limit) are: ($a$) Resolution of the Ca~{\sc ii}~H images is somewhat larger (e.g., $0.2~arcsec$) than the best resolution reached by {\sc Sunrise}. ($b$) The smaller features are not bright enough to be identified by our detection procedure. ($c$) The smaller features are not long-lived enough to be considered as a BP based on our restricted criteria (see Sect.~\ref{sec-analysis}).\\
Comparing the diameters obtained here with those reported in the literature is not straight forward, since they depend on the spatial resolution of the employed observations, the level of solar activity in the observed area (internetwork, network, plage, etc.) and on the technique used to determine the size. In this study, in order to avoid the effect of fine structure, we have set an upper limit on the size, so that we do not include the larger features that are also present in the {\sc Sunrise} data. We review the reported values of the BP sizes observed in the photosphere and the low chromosphere in Table~\ref{table:comp}. However, it is important to remember that the magnetic field of magnetic elements expands with height. Therefore, under the assumption that the brightness structures scale with the size of the magnetic features, a BP seen in the Ca~{\sc ii}~H passband should be larger than in the $G$-band, and larger still than in the TiO band (which corresponds essentially to the continuum outside sunspot umbrae) under the same observing conditions.\\
Using the network flux-tube model of \citet{Solanki1986} and the thin-tube approximation \citep{Defouw1976} we estimate the ratios of the flux tube diameter at the different heights of the atmosphere sampled by the red continuum (TiO band), the $G$-band (roughly $50~km$ higher;~\citealt{Carlsson2004}) and Ca~{\sc ii}~H ($\approx 440~km$ higher; see Sect.~\ref{subsec-CaIIHimages}). We obtain $1:1.1:2.6$, respectively, assuming the approximate mean formation heights. However, these ratios represent a lower limit of the size ratios, since the Ca~{\sc ii}~H BPs form in higher layers due to the effect of higher H$_{2V}$/H$_{2R}$ emission peaks around the Ca~{\sc ii}~H line-core (see Sect.~\ref{subsec-CaIIHimages}). One should keep these expansion rates in mind when comparing the size of BPs in different layers. Note that the Wilson depression affects all layers and does not, to first order, affect this ratio.\\
\citet{Berger1995} reported a range of FWHM between $0.17$ and $0.69~arcsec$, with an average value of $0.35~arcsec$ for photospheric $G$-band BPs. Using the same telescope and the same passband, \citet{Berger2001} noted a much smaller mean value of $0.145~arcsec$. This latter value is well bellow the theoretical diffraction limit, $0.2~arcsec$, of the telescope they used ($SVST$; Swedish Vacuum Solar Telescope) and hence smaller than the spatial resolution of their observations. This diameter probably reflects the threshold they used to obtain the size. In any case it suggests that they probably did not resolve most of the BPs in their sample. Later, \citet{Sanchez-Almeida2004} obtained a diameter of $0.25~arcsec$ on average (for GBPs), in data from a telescope with a higher diffraction-limit. \citet{Crockett2010} employed one-dimensional intensity profiling to detect magnetic GBPs and found a diameter distribution peaked at $0.31~arcsec$ which is in agreement with that of \citet{Berger1995}. Their largest GBPs were $1~arcsec$ in diameter. Recently, \citet{Riethmuller2012} studied the photospheric BPs from the {\sc Sunrise}/SuFI CN $3880~\AA$ passband and found an average diameter of $0.45~arcsec$ with a distribution of sizes ranging from $0.24$ to $0.78~arcsec$. A relatively small mean value of $0.16~arcsec$ was reported for internetwork BPs observed in high spatial resolution images taken in the TiO band~\citep{Abramenko2010}. An excessively large value of the diameter has been reported for the magnetic Ca~{\sc ii}~K BPs, observed in internetwork areas~\citep{Soltau1993}. He reported a diameter of $2.5~arcsec$ and considered it as an indication of a preferred size for the magnetic features. Obviously, either he considered very different features than we did or his spatial resolution was very low.\\
Based on the estimated expansion rates of a thin flux tube in the lower solar atmosphere, a photospheric GBP, i.e., the cross-sectional area of the flux tube in the height sampled by $G$-band, with a diameter of $0.35~arcsec$ (i.e., the largest mean value of a GBP's diameter reviewed here) expands to a diameter of $0.8-1.3~arcsec$ at the height sampled by Ca~{\sc ii}~H passband, assuming nearly vertical flux tubes. Such an expansion with height may lead to Ca~{\sc ii} H/K BPs with diameters as large as $2.5~arcsec$ \citep{Soltau1993} corresponding to relatively large GBPs, which, although rare, are observed~\citep[e.g.,][]{Berger1995,Crockett2010}.

\subsubsection{Lifetime}
\label{subsec-lifetimeHist}

Histogram of the lifetime for the $47$ BPs for which both birth and death times were observable are shown in Fig.~\ref{fig:stats}$d$. The red, dot-dashed line is an exponential fit with e-folding time of $526~sec$, roughly $9~min$. The condition that only BPs with lifetimes longer than $80~sec$ are considered in this study, in principle limits the left side of the distribution. As mentioned in Sect.~\ref{sec-analysis}, this lifetime threshold of $80~sec$ was introduced to exclude point-like brightenings caused by superpositions of waves and reversed granulation~\citep{de-Wijn2005}. However, the shortest lifetime we found is $167~sec$, so that in practice this limit should not influence the deduced lifetimes. This result suggests that either there is a lower limit to the lifetime of magnetic BPs, or one of the other criteria by which we isolate our BPs also discarded shorter lived BPs.\\
The longest lived BP in our sample lasted $1507~sec$. This is only slightly shorter than the longest image sequence we have used in this study and therefore, the length of our image series restricts the longest lifetime we can determine. This causes the number of longer lived BPs to be underestimated. Therefore, following~\citet{Danilovic2010}, we correct the lifetime distribution by multiplying it with a weight of

\begin{equation}
	\frac{\left ( n-2 \right )}{\left ( n-1-m \right )}\,,
	\label{equ:lifetime}
\end{equation}

\noindent
where $m$ is the number of frames that the BP lives and $n$ is the total number of frames in the observed time series (shown in shaded grey in Fig.~\ref{fig:stats}$d$). This results in a somewhat higher mean lifetime of $673~sec$, i.e., roughly $11~min$.\\
In an earlier study,~\citet{de-Wijn2005} discussed that tracking the Ca~{\sc ii}~H BPs in internetwork regions is harder than in the network due to the larger, more dynamic granules in the internetwork regions that cause crashing into the flux tubes. They concluded that this interaction can disturb the processes that cause the brightness of BPs and hence make them invisible. They reported a mean lifetime of $258~sec$ for their internetwork Ca~{\sc ii}~H BPs. The mean lifetime of $673~sec$ for internetwork Ca~{\sc ii}~H BPs that we have obtained from the {\sc Sunrise} data with a higher resolution, is a factor of $2.6$ larger than their value and is indicated by a vertical line in the lifetime histogram illustrated in Fig.~\ref{fig:stats}$d$. At least a part of this difference could be due to the effect of variable seeing on the results of~\citet{de-Wijn2005}, which can lead to significant underestimates of the lifetime.\\
\citet{Abramenko2010} showed that the majority ($\approx98.6\%$) of photospheric BPs, identified in high resolution images, have a lifetime less than $120~sec$. This implies either a fundamental difference between photospheric and chromospheric BPs. Alternatively, it is due to the difficulty of reliably identifying internetwork BPs in the visible continuum due to the generally low contrasts, particularly in the absence of polarisation information. Alternatively, it may be the effect of variable seeing in the observations of~\citet{Abramenko2010}. \citet{Nisenson2003} found an averaged lifetime of $9.2~min$ for their isolated, network GBP, which is comparable with our finding for isolated internetwork Ca~{\sc ii}~H BPs. \citet{de-Wijn2008} reported a mean lifetime of $10~min$ for the magnetic internetwork elements they studied in the quiet-sun. Although this is in a good agreement with what we obtained here, only $20$\% of their magnetic internetwork elements are related to BPs. They do not say if these BPs differ from the remaining magnetic features. In a lower resolution observation ($\approx0.3~arcsec$),~\citet{Muller1983} found a mean value of $1080~sec$ for network BPs and an average lifetime of $540~sec$ for internetwork BPs, both observed in the photospheric $5750~\AA$ passband. In addition, they also found a range of $180-600~sec$ for the lifetime of Ca~{\sc ii}~K BPs. Although at lower spatial resolution and affected by variable seeing, the result for the internetwork BPs is in agreement with our findings. Note that they also restricted themselves to the study of ``point-like'' BPs by setting an upper limit of $0.5~arcsec$ for the BPs' diameters.\\
Except for the results reported by~\citet{Muller1983} and \citet{Nisenson2003}, the lifetimes we find for internetwork BPs are significantly larger than those in the literature, both in the photosphere and the chromosphere. This may well be due to the stable conditions provided by {\sc Sunrise}. An alternative explanation for the longer lifetimes obtained in the {\sc Sunrise} data could be that through our criteria that allow a BP to disappear and appear again, we are sometimes falsely assigning two or more features to the same BP in the course of its lifetime. However, it has only a small effect on our lifetime distribution, as we found by recalculating the distribution without applying this criterion (only the few persistent flashers are affected). To test this situation, we consider the magnetic polarity of each BP for which this information is available. We find that every BP always maintains the same polarity during its lifetime. This supports our identifications. Our criterion to restrict our BPs in size may conceivably lead to a bias in lifetimes (if smaller BPs are longer lived). The agreement between our result and the one found by~\citet{Muller1983}, who also set an upper limit on BP sizes, may support such an interpretation.

\subsubsection{Circular Polarisation}
\label{subsec-CP}

\begin{figure}[h!]
	\centering
	\includegraphics[width=8cm]{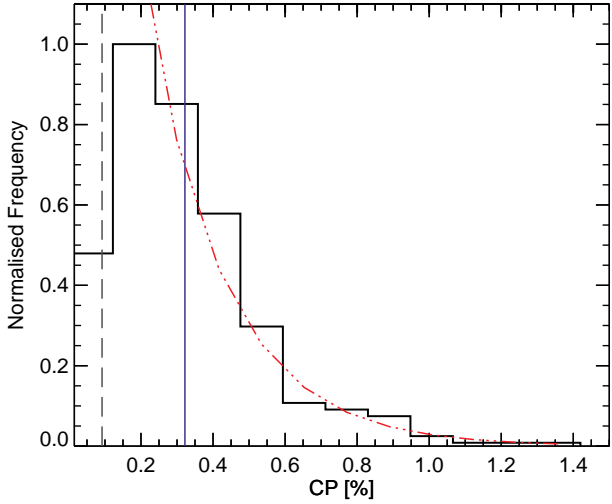}
	\caption{Histogram of the $CP$ at the corresponding positions of the {\sc Sunrise} Ca~{\sc ii}~H BPs in the photosphere. The dashed line marks the $3\sigma$ noise level. The red, dot-dashed line corresponds to an exponential fit and the vertical solid-line indicates a mean values of $0.32\%~CP$.}
	\label{fig:stats_V}
\end{figure}

Plotted in Fig.~\ref{fig:stats_V} is the distribution of all (unsigned) $CP$ values (described as in Sect.~\ref{subsec-stokes}) measured in individual snapshots of all $53$ BPs for which simultaneous IMaX observations were available (see Table~\ref{table:obslog}). The $3\sigma$ noise level is indicated by the vertical dashed line. The exponential fit to the distribution (restricted to $CP\geq0.2\%$) is overlaid (red dot-dashed line). It has an e-folding width of $0.21~CP~[\%]$. The vertical solid-line indicates the mean value of $CP$, which corresponds to $0.32\%$.\\
Our mean value of $CP$, from the spatially smoothed non-reconstructed IMaX data, is smaller, by a factor of about $8$, than that of~\citet{Riethmuller2012}, who measured this quantity at the position of their BPs from the phase-diversity reconstructed IMaX data. Hence, their larger photospheric CN BPs with diameters between $0.24-0.78~arcsec$ show much stronger polarisation signals compared to our small-scale, low chromospheric elements with diameters of $0.11-0.27~arcsec$. Note that due to the flux tube's expansion with height, our small Ca~{\sc ii}~H BPs are considered to have an even smaller size at the height sampled by the photospheric CN passband. Our measured $CP$ values (at the position of small BPs) from non-reconstructed data are roughly a factor of $2$ smaller than those computed from the phase-diversity reconstructed images.

\subsubsection{Longitudinal Magnetic Field}
\label{subsec-BcosGamma}

\begin{figure}[h!]
	\centering
	\includegraphics[width=8cm]{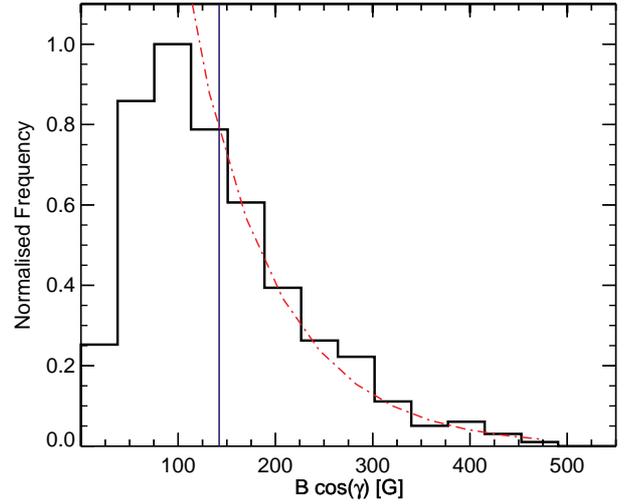}
	\caption{Histogram of the unsigned longitudinal component of magnetic field, $\vert B \cos(\gamma)\vert$, from the SIR inversion. The red, dot-dashed line represents an exponential fit to the right wing of the distribution, i.e., for $B>100~G$. The vertical line indicates the mean field strength value.}
	\label{fig:stats_b}
\end{figure}

As a final result, we perform an analysis of the Stokes profiles using the SIR code to compute the unsigned longitudinal component of magnetic field, $\vert B \cos(\gamma)\vert$ (see Sect.~\ref{subsec-inversion}). We show $\vert B \cos(\gamma)\vert$ since this value is retrieved more reliably than $B$, whose values can be adversely affected by noise in Stokes $Q$ and $U$. However, we should keep in mind that the $\vert B \cos(\gamma)\vert$ values may not be completely reliable, since the BPs either are not fully spatially resolved or the polarisation signals of the small BPs under study are too close to the noise level. The former issue causes the inversion code to underestimate the true field strength. The second issue is caused by the sometimes extreme line weakening of the highly temperature and magnetically sensitive IMaX Fe I $5250.2~\AA$ line~\citep{Sheeley1967,Stenflo1975,Shelyag2007,Lagg2010,Riethmuller2010,Riethmuller2012}. The fact that such point-like magnetic features are bright, suggests that they are the cross-sections of kG magnetic elements whose field strengths are underestimated, very likely because their true diameters in the photosphere are below the {\sc Sunrise} spatial resolution. A more thorough discussion is provided by~\citet{Riethmuller2012}.\\
Fig.~\ref{fig:stats_b} depicts the distribution of $\vert B \cos(\gamma)\vert$ for the BPs observed on 9 June 2009. This histogram shows a mean value of $142~G$ indicated by a vertical line in Fig.~\ref{fig:stats_b}. An exponential fit to the points with $B>100~G$ results in an e-folding width of $87~G$.\\
A similar conclusion as for the $CP$ distribution is made when our longitudinal magnetic field is compared with that of~\citet{Riethmuller2012}, who found stronger fields ranged between $25-1750~G$ and with a mean value of $544~G$. 

\subsection{A Relationship Between Intensity and Horizontal Velocity}

\begin{figure}[ht]
	\centering
	\includegraphics[width=8.5cm]{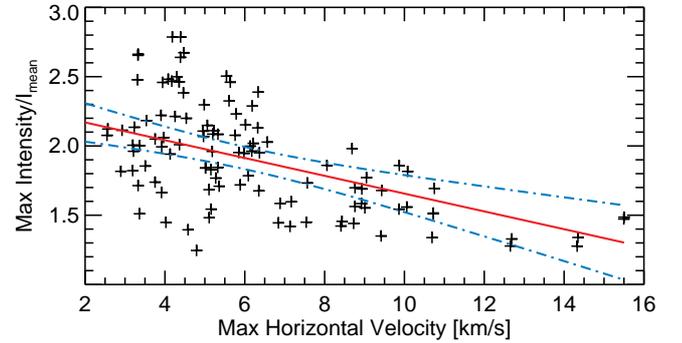}
	\caption{Relationship between the maximum horizontal velocity and the maximum intensity of the identified BPs. The solid (red) line illustrates a linear regression fit to the data with $95\%$ confidence bands (blue dot-dashed lines) around it.
              }
	\label{fig:vel_int}
\end{figure}

The only statistically significant correlation obtained from data of all BPs was found between the maximum values of intensity and maximum proper motion velocity of all identified BPs.\\
The maximum intensity values of all identified BPs are plotted versus their maximum horizontal velocities in Fig.~\ref{fig:vel_int}. The linear regression fit to the data (red solid line) shows an inverse correlation between maximum intensity and maximum proper motion velocity of the BPs with a correlation coefficient of $-0.49$. The blue dot-dashed lines in this plot indicate the confidence bands with $95\%$ confidence level which show an interval estimate for the entire regression line. Clearly, the faster a BP moves, the less bright it is. In particular, the combination of very bright and very mobile BPs is not present.

\section{Kink Wave Excitation Due to High-velocity Excursions}
\label{sec-kink}

\begin{figure}[ht]
	\centering
	\includegraphics[width=8.5cm]{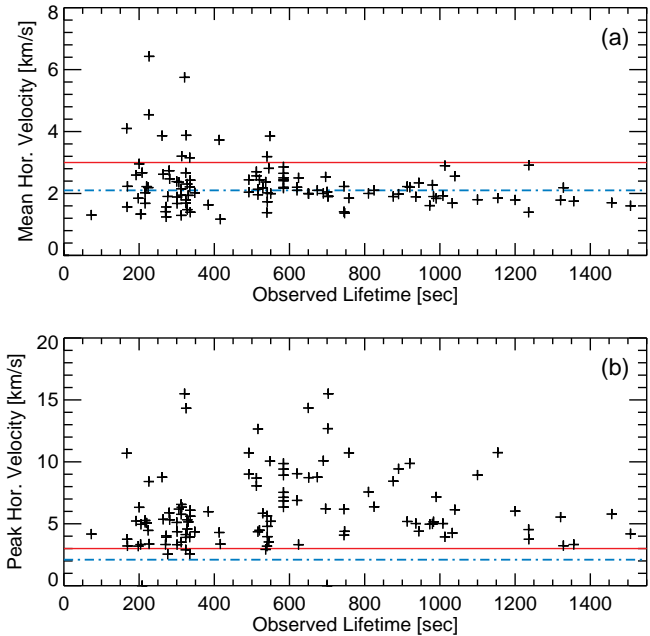}
	\caption{($a$) The mean horizontal velocity, and ($b$) maximum (peak) horizontal velocity versus the observed lifetimes of the studied BPs. The horizontal solid (red) and dot-dashed (blue) lines indicate the approximate limits for fast and intermediate BP velocities, respectively (see main text).
              }
	\label{fig:vel_life}
\end{figure}

We assume that the migration of our Ca~{\sc ii}~H BPs is a proxy for the motion of the cross-section of the underlying magnetic elements or flux tubes in the layer sampled by the {\sc Sunrise}/SuFI Ca~{\sc ii}~H passband, i.e., the cross-sections of magnetic elements at around the height of the temperature minimum/low chromosphere.\\
Furthermore, we found that all studied BPs have velocity variations similar to the one plotted in Fig.~\ref{fig:bp_var}$b$. In particular, the maximum (peak) velocity observed for each BP is often much larger than its mean value and is reached for only a brief time. This corresponds to the occurrence of intermittent pulses in the horizontal velocity. In Fig.~\ref{fig:vel_life} the mean and the peak values of horizontal velocity of all identified BPs are plotted versus their observed lifetimes. In both plots, higher velocities are mostly observed for shorter lived BPs. The peak horizontal velocity ranges between $2.6$ and $15.5~km~s^{-1}$, with a mean value of $6.2~km~s^{-1}$ for all BPs. In addition, Fig.~\ref{fig:vel_life}$b$ shows that almost all BPs move fast at least once during their lifetimes. The horizontal solid (red) and dot-dashed (blue) lines in both panels in Fig.~\ref{fig:vel_life} indicate the rough limits distinguishing fast, intermediate and slow BP motions, with fast motions implying mean proper motion speed $\gse3~km~s^{-1}$ and intermediate motions between $2$ and $3~km~s^{-1}$. These limits were chosen following \citet{Choudhuri1993a} and \citet{Kalkofen1997}.

\noindent
Note that vibration-induced image jitter introduced by instrumental effects (i.e., due to brief loss of pointing-lock of the telescope) into the {\sc Sunrise} observations is much smaller than our observed jerky motions~\citep{Berkefeld2011}.

\noindent
Rapid horizontal motions are of interest in particular if they occur as a pulse, i.e., in combination with a rapid acceleration, since such motions have been proposed to excite magnetoacoustic kink waves along flux tubes~\citep{Spruit1981,Choudhuri1993a}. Such transversal kink waves transport energy and can travel into the upper chromosphere without significant reflection~\citep{Spruit1981}. Although the reflection at the transition layer decreases the energy flux transported by the kink wave, it may be interesting for quiet coronal heating~\citep{Choudhuri1993b}. These early results have been given a strong boost by results of \citet{DePontieu2007} that transverse magneto-hydrodynamic (MHD) waves along spicules, which may propagate along extensions of BPs to higher layers, carry sufficient energy to heat the corona. According to \citet{Choudhuri1993a}, proper motions of BPs faster than $\approx2-3~km~s^{-1}$ (happening within a time shorter than $\approx100~sec$) are particularly effective in exciting kink waves.\\
This mechanism can be imagined as the rapid jostling of flux tube by granules in the photosphere which impulsively excites oscillations in the tube~\citep{Hasan2000}. Then, the oscillations (waves) propagate upwards while their velocity amplitudes increase exponentially until they reach values comparable to the tube speed for kink waves in chromospheric layers, where the kink waves become nonlinear~\citep[e.g.,][]{Kalkofen1997}. In the middle of the chromosphere, these transversal kink waves may couple to the longitudinal magnetoacoustic waves~\citep{Ulmschneider1991} and dissipate by forming shocks~\citep[e.g.,][]{Zhugzhda1995}.\\
\citet{Choudhuri1993a,Choudhuri1993b} carried out a theoretical analysis of the transversal kink waves induced by the flux tubes footpoints. They derived the following expression that gives a rough estimate of total energy flux ($F_{E}$) carried by the kink waves:

\begin{eqnarray}
F_{E}\,\approx\,6.5\times10^{26}\,\left ( \frac{\rho_{0}}{10^{-7}\, g\, cm^{-3}} \right )\left ( \frac{A_{0}}{10^{5}\, km^{2}} \right )\left ( \frac{v_{0}}{1\, km\, s^{-1}} \right )^{2} \nonumber\\
\cdot \left ( \frac{H}{250\, km} \right )\left ( \frac{F(\lambda)}{\lambda^{2} } \right )n\, f \; \; \; ergs\, cm^{-2}\, s^{-1}\,,
	\label{equ:energy_flux}
\end{eqnarray}	

\noindent
where $\rho_{0}$ is the atmospheric density, $A_{0}$ is the cross sectional area of the flux tube (i.e., area of the BP), $v_{0}$ is the maximum velocity of a horizontal pulse, and $H$ is the scale height. The parameters $n$ and $f$ in Eq.~(\ref{equ:energy_flux}) represent the number density of BPs and the frequency of motions (i.e., pulses), respectively. $\lambda$ is a dimensionless parameter related to the rapidity of a BP, i.e., a measure of the strength of the BP motion. This parameter is determined by $\lambda=v_{0}/(\omega_{c} L)$, where $\omega_{c}$ is the cut off frequency at a certain layer in the atmosphere and $L$ is the BP displacement within the pulse duration.
The dimensionless function $F(\lambda)$ is the asymptotic energy which depends on the employed atmospheric model. \citet{Choudhuri1993b} examined three different models and finally concluded that an isothermal atmosphere is a reasonable model for kink waves propagation in the solar atmosphere induced by rapid flux tube footpoint motions. Therefore, we employ the isothermal atmosphere model in our calculations.

\begin{table*}[ht!]
\caption{Characteristics$^{\mathrm{ *}}$ of Ca~{\sc ii}~H BPs pulses and estimated values of energy flux for different velocity bins.}             
\label{table:energyFlux}      
\centering                        
\begin{tabular}{l c c c c c c c c c}     
\hline\hline                
Velocity Bin & $N_{p}$ & $v_{0}$ & $t_{p}$ & $A_{0}$ & $L$ & $f$ & $\lambda$ & $F(\lambda)$ & $F_{E}$\\    
 $[km~s^{-1}]$ &  & [$km~s^{-1}$] & [$s$] & [$km^{2}$] & [$km$] & [$mHz$] &  &  & [$erg~cm^{-2}s^{-1}$]\\  
\hline                        
 $2-3$     & 252 & 2.5 & 42 & 16000 & 55 & 4.00 & 1.5 & 0.33 & $4.1\times10^4$\\
 $3-4$     & 206 & 3.5 & 43 & 16500 & 61 & 3.00 & 1.9 & 0.48 & $5.9\times10^4$\\
 $4-5$     & 135 & 4.4 & 43 & 17100 & 70 & 2.00 & 2.1 & 0.54 & $6.1\times10^4$\\
 $5-6$     & 84 & 5.4 & 47 & 17600 & 70 & 1.00 & 2.6 & 0.70 & $5.0\times10^4$\\
 $6-7$     & 31 & 6.4 & 42 & 16900 & 81 & 0.50 & 2.6 & 0.71 & $2.4\times10^4$\\
 $7-8$     & 15 & 7.5 & 33 & 18300 & 72 & 0.20 & 3.4 & 0.98 & $1.4\times10^4$\\
 $8-9$     & 20 & 8.4 & 38 & 18600 & 80 & 0.30 & 3.5 & 1.00 & $2.4\times10^4$\\
 $9-10$   & 11 & 9.5 & 40 & 18600 & 74 & 0.20 & 4.3 & 1.25 & $1.4\times10^4$\\
 $10-11$ & 7 & 10.4 & 22 & 18600 & 74 & 0.10 & 4.7 & 1.39 & $1.0\times10^4$\\
 $11-12$ & 2 & 11.4 & 47 & 25000 & 46 & 0.03 & 8.2 & 2.58 & $2.7\times10^3$\\
 $12-13$ & 2 & 12.4 & 28 & 19800 & 70 & 0.03 & 5.9 & 1.81 & $3.5\times10^3$\\
 $14-15$ & 3 & 14.2 & 49 & 19700 & 76 & 0.04 & 6.2 & 1.90 & $6.5\times10^3$\\
 $15-16$ & 2 & 15.7 & 33 & 22300 & 65 & 0.03 & 8.0 & 2.49 & $4.8\times10^3$\\
\hline                                   
 $2-16$ & 770 & 4.1 & 42 & 16800 & 63 & - & - & - & $3.1\times10^5$\\
\hline                                 
\end{tabular}
\begin{list}{}{}
\item[$^{\mathrm{*}}$] $N_{p}$: Number of pulses; $v_{0}$: Mean horizontal velocity; $t_{p}$: Mean lifetime; $A_{0}$: Mean area; $L$: Mean displacement; $f$: Frequency of pulse occurrence; $\lambda$: Rapidity parameter; $F(\lambda)$: Asymptotic energy; $F_{E}$: Net flux
\end{list}
\end{table*}
\noindent
Since we are aiming for a rough estimate of the energy flux generated by our high-velocity Ca~{\sc ii}~H BPs, we are satisfied with approximate values of the various parameters entering Eq.~(\ref{equ:energy_flux}). Following~\citet{Wellstein1998}, we take the scale height $H$ in an isothermal atmosphere of $7000~K$ (as an upper limit), equal to $\approx150~km$. At the height of formation of the {\sc Sunrise}/SuFI Ca~{\sc ii}~H BPs (i.e., a height corresponding roughly to the temperature minimum), we use the density from the VAL-C atmospheric model~\citep{Vernazza1981}: $\rho_{0}\approx6\times10^{-9}~g~cm^{-3}$.\\
For the acoustic cut off frequency at the temperature minimum we use $3\times10^{-2}~s^{-1}$~\citep{Spruit1981}.\\
Then, we analyse the horizontal velocity variations of each BP and look for significant pulses, i.e., when the proper motion velocity exceeds a certain limit for a maximum duration of $100~sec$. We compute the pulses occurring in $1~km~s^{-1}$ wide velocity bins between values of $2$ and $16~km~s^{-1}$. The pulse parameters, i.e., the peak horizontal velocity and the area of each BP at that time, as well as the displacement which the BP suffers during each pulse, are recorded. The frequency of pulses for each bin is then computed as the ratio of the number of pulses in the bin and the integrated lifetime over all BPs. As mentioned earlier, the identified BPs in this study have a number density of $0.03~(Mm)^{-2}$.\\
The parameters $v_{0}$, $A_{0}$ and $L$ are computed as the average over all the peak horizontal velocities, all the areas and all the displacements determined for all pulses in all BPs within each velocity bin, respectively. We extrapolate the plot of $F(\lambda)$ versus $\lambda$ from Fig.~6 in~\citet{Choudhuri1993a} to obtain the value of $F(\lambda)$ for the computed $\lambda$s. The values of the obtained parameters for each velocity bin are presented in Table~\ref{table:energyFlux}. The integrated energy flux is then computed over all velocity bins larger than a given speed. We do the analysis for two speed limits: one for $v=3~km~s^{-1}$ and one for $v=2~km~s^{-1}$ representing fast and intermediate BP motions, respectively~\citep{Choudhuri1993a,Kalkofen1997}.\\
We find an integrated value of $F_{E}\approx2.7\times10^{5}~erg~cm^{-2}s^{-1}$ (or $\approx270~W~m^{-2}$) for pulses with $v\ga3~km~s^{-1}$ and a marginally larger value of $F_{E}\approx3.1\times10^{5}~erg~cm^{-2}s^{-1}$ (or $\approx310~W~m^{-2}$) for pulses with $v\ga2~km~s^{-1}$. These estimated values of total energy flux are compatible with the estimated energy flux density of $1-3\times10^{5}~ergs~cm^{-2}s^{-1}$ ($100-300~W~m^{-2}$) required to heat the quiet corona and/or drive the solar wind~\citep{Hollweg1990,Hansteen1995}. Our value is a lower limit since we only considered small, internetwork magnetic elements.\\
Table~\ref{table:energyFlux} is in agreement with the finding of Choudhuri et al. (1993b), that faster motions, though more infrequent, provide more substantial contributions compared to the slower ones (cf. Eq.~\ref{equ:energy_flux}).\\
Note that if the formation height of the Ca~{\sc ii}~H channel were to lie lower, e.g., in the mid photosphere where the atmospheric density is much higher, $\rho_{0}\approx5\times10^{-8}~g~cm^{-3}$, then the energy flux estimated based on Eq.~(\ref{equ:energy_flux}) would be about one order of magnitude larger.\\
\citet{Muller1994} reported a very much larger value of $F_{E}=2000~W~m^{-2}$ for their photospheric network BPs with $v>2~km~s^{-1}$ whereas \citet{Wellstein1998} found a value of $F_{E}=440~W~m^{-2}$ for their K$_{2V}$ BPs with mean horizontal velocity of $6~km~s^{-1}$. The latter value is highly affected by the formation height assumption of $1100~km$ for their Ca~{\sc ii}~K passband, and may not be relevant in any case since it is unclear which fraction of their features really is magnetic. The results of \citet{Muller1994} suggest that it would be worthwhile to follow also larger BPs as found in the network in {\sc Sunrise} data.

\section{Conclusions}
\label{sec-conclusions}

In this paper, we have provided observational properties of intrinsically magnetic and highly dynamic small bright point-like features in the {\sc Sunrise} Ca~{\sc ii}~H passband. BPs of the type studied here, i.e., with diameters below $0.3~arcsec$, are present mainly in the internetwork. We have, for simplicity, further restricted our analysis to those BPs that do not merge or split in the course of their observed lifetime. They cover only 5\% of the Sun's surface and radiate on average $1.48$ times more than the average quiet-Sun in the $1.8~\AA$ broad {\sc Sunrise} Ca~{\sc ii}~H filter, so that $7.5$\% of the radiative energy losses in lower chromospheric layers are due to such features (assuming Ca~{\sc ii}~H to be representative of other spectral lines formed in that layer). This is negligible. However, motions of these features may be responsible for kink waves travelling into the corona, and for the braiding of field lines leading to microflares~\citep{Parker1988}, so that studying them can potentially help us in our understanding of chromospheric and possibly coronal heating mechanisms.\\
We applied a set of stringent criteria in order to separate the (magnetic) BPs we wish to study from all other bright, point-like features. In particular, we only consider BPs smaller than $0.3~arcsec$.\\
This limited us to a smaller number of BPs, but ensured that they are not oscillations or wave-like features, nor the product of reversed granulation.\\
Nearly all the BPs for which we could test this were found to be associated with significant photospheric magnetic field.
This result agrees with that of~\citet{de-Wijn2005}, who showed, although at lower spatial resolution, that internetwork Ca~{\sc ii}~H BPs were located inside magnetic patches. They also found a good spatial coincidence between their Ca~{\sc ii}~H BP patches and the ones they found in $G$-band images. This is also the case in the {\sc Sunrise} data~\citep{Riethmuller2010}.\\
These small magnetic BPs in internetwork areas have long lifetimes with a mean value of $673~sec$, much longer than any values in the literature for Ca~{\sc ii}~H BPs, but in good agreement with the mean lifetime of about $9~min$ for photospheric internetwork BPs~\citep{Muller1983} as well as a mean lifetime of $10~min$ for the internetwork magnetic elements (IME;~\citealt{de-Wijn2008}), of which only $20\%$ are associated with $G$-band and Ca~{\sc ii}~H BPs. Our internetwork features appear to live less long than (photospheric) network BPs which last on average $1080~sec$~\citep{Muller1983}, however, a comparable mean lifetime of $9.2~min$ for network GBPs was found by~\citet{Nisenson2003}.
In addition, we showed that the Ca~{\sc ii}~H BPs move horizontally with an average speed of $2.2~km~s^{-1}$. We suspect that the much higher horizontal velocities reported by \citet{Wellstein1998} and \citet{Steffens1996} refer to a different type of feature, which is probably non-magnetic. Furthermore, an inverse correlation between the maximum values of intensity and horizontal velocity of the identified BPs was observed, i.e., the BPs are brightest when they are at rest.\\
It was shown that in addition to the considerable mean values of the horizontal velocity of the subsonic BPs, almost all {\sc Sunrise} Ca~{\sc ii}~H BPs move with a much larger peak horizontal velocity (up to $15.5~km~s^{-1}$) for at least a short period of time. On average a BP spends $3.5\%$ of its life travelling at a supersonic speed. These motions are not regular but often correspond to pulses of rapid horizontal motion.
Such jerky, pulse-like motions efficiently give rise to kink waves travelling along a magnetic flux tube according to~\citet{Choudhuri1993a,Choudhuri1993b}. They calculated that the efficiency of kink wave excitation depends on the rapidity of the BP motion with respect to the cutoff frequency of the atmosphere. We computed a rough estimate of the energy flux generated by the rapid BP motions (i.e., for pulses with $v\ga2~km~s^{-1}$) of $\approx3.1\times10^{5}~erg~cm^{-2}s^{-1}$ (or $\approx310~W~m^{-2}$), which is sufficient to heat the quiet corona (energy flux density of $1-3\times10^{5}~ergs~cm^{-2}s^{-1}$ or $100-300~W~m^{-2}$;~\citealt{Withbroe1977,Hollweg1990,Hansteen1995}) if the kink pulses propagate into the corona and dissipate their energy there. Our estimated value is an order of magnitude lower than the energy flux transported by Alfv\'{e}n waves estimated by~\citet{DePontieu2007} from observations of type~{\sc ii} spicules ($4$ to $7~kW~m^{-2}$). However, the mechanisms by which these different energy fluxes are generated may be different. \citet{Hasan2008} argued that the heating in Ca~{\sc ii}~H BPs, caused by weak shocks occurring at short time intervals (less than $100~sec$) in magnetic flux tubes, is in contrast with the long-period waves which are considered as the spicules driver~\citep{DePontieu2004}. Also, by restricting ourselves to small BPs, we may be missing the magnetic features carrying the most energy into the upper atmosphere.\\
The mean value of the unsigned longitudinal component of magnetic field, $\vert B \cos(\gamma)\vert$, in the BPs was found to be $142~G$, reaching $500~G$ only rarely. These values may be influenced by some of the BPs not being fully spatially resolved \citep{Riethmuller2012} and/or the polarisation signals lying close to the noise level at some times, as can be seen in Fig.~\ref{fig:stats_V}.\\
All distributions of the parameters studied here, except that of diameter, display an exponential fall-off towards larger values, implying random stochastic processes. The shape of the size histogram is partly determined by our restriction to features smaller than $0.3~arcsec$, and may well be affected by the fact that many of these features could be unresolved in the photosphere, as is suggested by their weak fields (see~\citealt{Riethmuller2012}).\\
We identified $7$ persistent flashers~\citep{Brandt1992} in our sample of BPs. We find that the flashers are normal BPs that differ mainly in that they have smaller average values of intensity, so that their brightness drops below the threshold at some moments in time and they seem to disappear temporarily.\\
We expect that granular and intergranular motions are primarily responsible for the horizontal motions of the corresponding flux tubes rooted in intergranular lanes. Interestingly, the lifetime distribution of our Ca~{\sc ii}~H BPs (range between $167$ and $1507~sec$) is comparable with that found by~\citet{Hirzberger1999} for granules, who showed a range of $168-1800~sec$ for granular lifetimes with only a relatively small number of long-lived granules with lifetimes longer than $1200~sec$. However, they found a mean lifetime of about $360~sec$ for the granules which is smaller than the lifetime of our Ca~{\sc ii}~H BPs. In addition, the existence of horizontal convective supersonic flows at the boundaries between granules and intergranular lanes~\citep[e.g.,][]{Cattaneo1989,Solanki1996,Nordlund2009,Bellot2009,Vitas2011} may explain our observations of fast motions BPs and supersonic pulses which can be due to the impact of fast granular flows on the flux tubes.\\
Summarising, we studied the structure and dynamics of the smallest currently observable magnetic bright features at the height sampled by the {\sc Sunrise}/SuFI Ca~{\sc ii}~H passband (i.e., a height corresponding roughly to the temperature minimum). We determined and discussed the statistical distributions of the BPs' properties including intensity, horizontal velocity, size, lifetime, polarisation signals as well as magnetic field strength. We found an anti-correlation between the maximum proper motion velocity and the peak intensity values of the BPs. With the help of an advanced image processing technique, selecting the actual magnetic BPs based on stringent criteria and using data unaffected by seeing from {\sc Sunrise}, we provided accurate horizontal velocity profiles of the Ca~{\sc ii}~H BPs. These profiles revealed indications of fast pulses (i.e., rapid BP motions on short time scales) which may contribute to coronal heating by exciting kink waves in the corresponding flux tubes. The energy flux that we estimate following \citet{Choudhuri1993a,Choudhuri1993b} is marginally sufficient to heat the quiet corona. Note that we restricted ourselves to small internetwork magnetic elements, so that we are probably including only a small fraction of the energy flux in our estimate. Hence, in addition to other possibilities of coronal heating, e.g., braiding of field lines leading to microflares~\citep{Parker1988}, the jerky motions of the BPs studied in this work, can excite waves that potentially carry enough energy to contribute to the heating of the quiet corona. Future work will aim at detecting such waves.

\begin{acknowledgements}

The German contribution to {\sc Sunrise} is funded by the Bundesministerium f\"{u}r Wirtschaft und Technologie through Deutsches Zentrum f\"{u}r Luft- und Raumfahrt e.V. (DLR), Grant No. 50 OU 0401, and by the Innovationsfond of the President of the Max Planck Society (MPG). The Spanish contribution has been funded by the Spanish MICINN under projects ESP2006- 13030-C06 and AYA2009-14105-C06 (including Euro- pean FEDER funds). The HAO contribution was partly funded through NASA grant NNX08AH38G. This work has been partly supported by the WCU grant (No. R31- 10016) funded by the Korean Ministry of Education, Science and Technology. We thank Han Uitenbroek for providing his RH-code and the model atmospheres. We also wish to thank the anonymous referee for valuable comments. SJ acknowledges a PhD fellowship of the International Max Planck Research School on Physical Processes in the Solar System and Beyond.

\end{acknowledgements}

\bibliographystyle{aa} 
\bibliography{ref_dynamics} 

\end{document}